\documentclass[]{iopart}

\expandafter\let\csname equation*\endcsname\relax
\expandafter\let\csname endequation*\endcsname\relax

\usepackage[unicode]{hyperref}
\usepackage[all]{hypcap}
\usepackage{amssymb,amsmath,verbatim,mathtools,needspace,enumitem,etoolbox,graphicx,microtype,pifont,url,cite,physics}
\usepackage[usenames,dvipsnames]{xcolor}
\usepackage[T1]{fontenc}
\usepackage[utf8]{inputenc}
\usepackage[normalem]{ulem}
\definecolor{linkcolor}{rgb}{0.0,0.3,0.5}
\hypersetup{colorlinks=true,linkcolor=linkcolor,citecolor=linkcolor,filecolor=linkcolor,urlcolor=linkcolor}
\usepackage{orcidlink}
\usepackage{subcaption}

\newcommand{\ethz}{{ETH Zurich, Department of Physics, Institute for Particle and Astrophysics,
Wolfgang-Pauli-Str. 27, 8093 Zurich, Switzerland}}
\newcommand{\milan}{{Dipartimento di Fisica ``G. Occhialini'', Universit\'a degli Studi di Milano-Bicocca, Piazza della Scienza 3, 20126 Milano, Italy}}
\newcommand{\infn}{{INFN, Sezione di Milano-Bicocca, Piazza della Scienza 3, 20126 Milano, Italy}}

\begin{document}

\begin{center}
\title[G.~Giarda et al.]{Accelerated inference of binary black-hole populations from the stochastic gravitational-wave background}
\end{center}

\author{
Giovanni Giarda$^{1,2}$ \orcidlink{0009-0007-3093-7821},
Arianna I. Renzini$^{2,3}$ \orcidlink{0000-0002-4589-3987},
\\ Costantino Pacilio$^{2,3}$ \orcidlink{0000-0002-8140-4992},
Davide Gerosa$^{2,3}$ \orcidlink{0000-0002-0933-3579}
}
\vspace{0.1cm}
\address{$^{1}$~\ethz}
\address{$^{2}$~\milan}
\address{$^{3}$~\infn}

\ead{\href{mailto:g.giarda2@campus.unimib.it}{\rm g.giarda2@campus.unimib.it}}

\setcounter{footnote}{0}

\begin{abstract}

Third-generation ground-based gravitational wave detectors are expected to observe $\mathcal{O}(10^5)$ of overlapping signals per year from a multitude of astrophysical sources that will be computationally challenging to resolve individually. 
On the other hand, the stochastic background resulting from the entire population of sources encodes information about the underlying population, allowing for population parameter inference independent and complementary to that obtained with individually resolved events. %
Parameter estimation in this case is still computationally challenging, as computing the power spectrum involves sampling $\sim 10^5$ sources for each set of hyperparameters describing the binary population. In this work, we build on recently developed importance sampling techniques to compute the SGWB efficiently and train neural networks to interpolate the resulting background. We show that a multi-layer perceptron can encode the model information, allowing for significantly faster inference. We test the network assuming an observing setup with CE and ET sensitivities, where for the first time we include the intrinsic variance of the SGWB in the inference, as in this setup it presents a dominant source of measurement noise.
\end{abstract}


\section{Introduction}\label{sec:introduction}

In the first ten years of gravitational-wave (GW) astronomy, the LIGO-Virgo-KAGRA (LVK) network has observed numerous signals from compact binary coalescences~\cite{2019PhRvX...9c1040A,2021PhRvX..11b1053A,2023PhRvX..13d1039A,2024PhRvD.109b2001A}.
Ongoing technological advances will bring about third-generation (3G) ground-based GW detectors, such as the Einstein Telescope \cite{2025arXiv250312263A} and Cosmic Explorer \cite{2019BAAS...51g..35R}.
These new 10 km-scale detectors are expected to detect $O(10^5)$ signals from compact binaries per year. %
Many of these will overlap in the data stream, and the computational challenges of resolving and characterizing each event individually will be extreme \cite{2024arXiv241202651H}. The cumulative signal sourced by the binary population is that of a stochastic gravitational-wave background (SGWB), which carries information about the underlying population features, and can be leveraged to perform population parameter inference ~\cite{2020ApJ...896L..32C}. To this end, the SGWB is usually combined with individual event detection. On the contrary, in this paper we propose a method to infer the binary population parameters solely from the stochastic signal, avoiding the need to infer parameters of each individual source first.

Evaluating the SGWB as a function of the population parameters is an expensive but unavoidable operation during inference. It typically involves sampling realizations of individual events and evaluating their corresponding waveforms, which is an inherently costly process. Recent methods leveraging importance sampling techniques \cite{2024A&A...691A.238R} have significantly accelerated this computation, yet they still overlook the intrinsic variance arising from the finite number of events used in the calculation.

In this work, we present further advances in this direction. In particular, we (i) accelerate the computation of the expected value of the SGWB using a neural network and (ii) include the associated variance during inference to improve robustness. 
Parameter estimation involving SGWB measurements with ground-based detectors~\cite{isotropic_O1,isotropic_O2, isotropic_O3,2020ApJ...896L..32C} has thus far ignored the effect of the signal's intrinsic variance, as currently measurements are entirely noise-dominated, and only produce upper limits of the stochastic signal. In this paper, we consider cases where this intrinsic variance dominates the measurement, lying well above the uncertainty due to detector noise, and consistently include it in the inference process for the first time.

This paper is organized as follows. In Sec.~\ref{sec:problem}, we introduce the SGWB computation problem; in Sec.~\ref{sec:population}, we describe the population models employed; in Sec.~\ref{sec:mlp}, we present our network training and results, and discuss the separate modeling of the SGWB variance; in Sec.~\ref{sec:inference}, we analyze the resulting performance in inference tasks; and in Sec.~\ref{sec:conclusion}, we draw conclusions and discuss potential future directions.

\section{Stochastic gravitational-wave background modeling}
\label{sec:problem}

\subsection{Relevant quantities}
The fractional energy density spectrum of gravitational waves is given by~\cite{2001astro.ph..8028P}:
\begin{equation}
    \Omega_{\mathrm{GW}}(f) = \frac{1}{\rho_c}\frac{\dd\rho_{\mathrm{GW}}(f)}{\dd\ln f}\,,
\end{equation}
where $\rho_\mathrm{GW}(f)$ is the GW energy density at a frequency $f$ and  $\rho_c = 3H_0^2/8\pi G$ is the critical energy density. 

Let us denote the properties of individual GW sources (e.g., masses, redshift) by $\mathbf{\Theta}$. For a finite number of sources $N_s$, the energy density spectrum $\Omega_{\mathrm{GW}}$ can be estimated as the average fractional GW energy density detected over an observation period $T_{\mathrm{obs}}$ \cite{Meacher:2015iua},
\begin{equation}
    \Omega_{\mathrm{GW}}(f) \simeq \frac{f^3}{T_{\mathrm{obs}}}\frac{4\pi^2}{3H_0^2} \sum^{N_s}_i \mathcal{P}_d (\mathbf{\Theta}_i;f)\,,
\end{equation}
where 
\begin{equation}
    \mathcal{P}_d (\mathbf{\Theta};f) = \tilde{h}^2_+(\mathbf{\Theta};f) + \tilde{h}^2_\times(\mathbf{\Theta};f)
\end{equation}
is the unpolarized power in the detector frame associated with a GW event with parameters $\mathbf{\Theta}$ in the frequency domain and $\tilde{h}_{+,\times}(\mathbf{\Theta},f)$ are the polarizations of the GW waveform, also in the frequency domain. %
A single realization of the $\Omega_{\mathrm{GW}}$ spectrum is determined by a particular set of source parameters $\mathbf{\Theta}_i$. In contrast, its ensemble average, denoted by $\overline{\Omega}_{\mathrm{GW}}$, is determined by the distribution of $\mathbf{\Theta}$, which is characterized through a set of population parameters $\mathbf{\Lambda}$. %
In the limit of infinite observation time and number of events, the observed spectrum converges to its ensemble average~\cite{2001astro.ph..8028P, 2024A&A...691A.238R,2020ApJ...896L..32C}: 
\begin{equation} \label{phinney_gwb}
    \overline{\Omega}_{\mathrm{GW}}(\mathbf{\Lambda}; f) = f^3 \frac{4\pi^2}{3H^2_0} R \int \dd\mathbf{\Theta} \, p_d(\mathbf{\Theta}|\mathbf{\Lambda}) \, \mathcal{P}_d(\mathbf{\Theta};f),
\end{equation}
where $p_d(\mathbf{\Theta}|\mathbf{\Lambda})$ is the (normalized) probability distribution for the GW parameters and $R$ is the total rate of events per unit detector-frame time. 

For a fixed population described by a set of parameters $\mathbf{\Lambda}$, individual realizations of $\Omega_{\mathrm{GW}}(f)$ will exhibit fluctuations around the expected value $\overline{\Omega}_{\mathrm{GW}}(f)$ due to the finite number of contributing sources. %
This variability, which we refer to hereinafter as \emph{intrinsic}, is linked to the underlying stochasticity of the background, and is particularly relevant when the number of sources contributing to the signal is small (e.g., for short observation times). Accounting for this intrinsic variance is essential in assessing inference procedures based on the SGWB spectrum. 

At this stage, it is important to note that, thus far, inference on the quantity $\overline{\Omega}_{\rm GW}$ from a population of stellar mass black holes has assumed that detectors are weakly sensitive to its intrinsic variance, and that measurements are dominated by detector noise~\cite{Romano:2016dpx}. %
This will no longer be the case for 3G ground-based detectors, as it is not the case for the SGWB observed by pulsar timing arrays~\cite{Lamb:2024gbh} and for stochastic signals potentially observed by LISA. %
In this paper, we include the intrinsic signal variance in the inference process for the first time.

\subsection{Importance sampling}

A naive Monte-Carlo evaluation of Eq.~(\ref{phinney_gwb}) with direct sampling from $p_d(\mathbf{\Theta}|\mathbf{\Lambda})$ is a prohibitive operation, even for straightforward parameter space explorations. Current approaches, such as the \textsc{popstock} implementation \cite{2024A&A...691A.238R}—which is built on top of \textsc{bilby} \cite{2019ApJS..241...27A} and \textsc{gwpopulation} \cite{2019PhRvD.100d3030T}—rely on importance sampling.

Importance sampling approximates the process of sampling from a target distribution by reweighting samples drawn from a proposal distribution. For the case of the SGWB, one can sample from a compact binary population described by population parameters $\mathbf{\Lambda}_0$, and then reweight in favour of the desired population $\mathbf{\Lambda}_i$. This may be written as \cite{2024A&A...691A.238R}:
\begin{align}
    \int \dd \mathbf{\Theta} \, p_d(\mathbf{\Theta}|\mathbf{\Lambda}_i) \mathcal{P}_d (\mathbf{\Theta}) 
    &= \int \dd\mathbf{\Theta} \,w_{i}(\mathbf{\Theta}) \, p_d(\mathbf{\Theta}|\mathbf{\Lambda}_0) \mathcal{P}_d (\mathbf{\Theta})\,.
   \label{iint}
\end{align}
where the weight 
\begin{equation}
    w_{i}(\mathbf{\Theta}) = \frac{p_d(\mathbf{\Theta}|\mathbf{\Lambda}_i)}{p_d(\mathbf{\Theta}|\mathbf{\Lambda}_0)} 
\end{equation}
is computed for each sample $\mathbf{\Theta}_j$ extracted from the fiducial population. Provided that a sufficiently large number of samples is used, one can then approximate the integral of Eq.~(\ref{iint}) with a Monte Carlo summation:
\begin{equation}
    \label{eq:MC:1}
    \int \dd\mathbf{\Theta}\, p_d(\mathbf{\Theta}|\mathbf{\Lambda}_i) \mathcal{P}_d (\mathbf{\Theta})  \simeq \sum_j w_{i} (\mathbf{\Theta}_j)\mathcal{P}_d (\mathbf{\Theta}_j)\,.
\end{equation}

This re-weighting methodology allows for significant computational efficiency because the GW waveforms required to compute $ \mathcal{P}_d (\boldsymbol\Theta_j) $ are evaluated once and used for several values of $\boldsymbol\Lambda_i$. A key aspect of importance sampling is controlling the variance of the weights \cite{2019RNAAS...3...66F}. A large variance in the weights can lead to inaccurate estimates of the target distribution, as the results become dominated by a few highly weighted samples. This typically occurs when the proposal distribution poorly matches the target distribution. One can quantify this mismatch using the effective sample size
\begin{equation}
    N_{\mathrm{eff}} = \frac{\left[ \sum_{j=1}^{N_{\mathrm{s}}} w_{i} (\mathbf{\Theta}_j) \right]^2}{\sum_{j=1}^{N_{\mathrm{s}}} w_{i}^2 (\mathbf{\Theta}_j)},
\end{equation}
where $N_{\mathrm{s}}$ is the total number of samples. In practice, a reliable estimate should have $N_{\mathrm{eff}} \gg 1$, and as close as possible to $N_{\mathrm{s}}$. In Ref.~\cite{2024A&A...691A.238R}, the re-weighted SGWB spectrum was validated by comparison with direct Monte Carlo estimates. They found that employing a number of sample sources of $N_s \geq 10^5$ for the spectral computation is sufficient for the results to lie within the intrinsic sample variance of the Monte Carlo realizations in all cases. %
In all the cases considered below, the median effective sample size is approximately $ N_{\mathrm{eff}} \simeq 2 \times 10^4 $, indicating a well-behaved weight distribution. 

\subsection{Neural networks}

While importance sampling significantly reduces the cost of exploring the parameter space, it remains a bottleneck in inference tasks that require repeated evaluations of $\overline{\Omega}_{\mathrm{GW}}$. This limitation is particularly evident in rapid parameter estimation with large sample sets $N_{\rm s}$, where reweighting (which requires calculating the probability of each sample) can become prohibitively time consuming, as well as when inferring 
complicated population models with many features, which are inherently hard to re-weight and lead to low $N_{\rm eff}$\footnote{We point out that, while in this paper we employ importance sampling to train the network, this is not strictly required, and one could take the (lengthier) route of direct Monte-Carlo evaluation of \ref{phinney_gwb} to train the network.}. 
Furthermore, the summation \eqref{eq:MC:1} is affected by an intrinsic variance due to the finite number of Monte Carlo samples, which is directly interpreted as a variance on ${\overline{\Omega}}_{\mathrm{GW}}$ due to the specific event realization used in the calculation. %
Directly modelling the ensemble average $\overline{\Omega}_{\mathrm{GW}}$, rather than a single realization, and including an estimate of its variance, will improve the robustness of the inference process by mitigating the impact of stochastic fluctuations inherent to the SGWB. %
To address this, we propose interpolating $\overline{\Omega}_{\mathrm{GW}}$ using an multi-layer perceptron (MLP), enabling faster and more stable evaluations, and explicitly accounting for the variance of $\Omega_{\rm GW}$ about $\overline{\Omega}_{\rm GW}$ when defining the likelihood for $\mathbf{\Lambda}$.

MLPs are feed-forward neural networks with an input layer, one or more hidden layers, and an output layer. They are universal function approximators \cite{1989MCSS....2..303C, hornik1989multilayer}, making them well-suited for interpolating complex functions, while remaining relatively simple to implement and train. A similar approach for the SGWB originating from supermassive black holes has been recently presented in Ref.~\cite{2024A&A...687A..42B}, relevant in the context of pulsar-timing array measurements. In Ref.~\cite{2024A&A...687A..42B} the background mean and variance are modeled together, although the network was not used for inference tasks.

For each set of hyperparameters $\mathbf{\Lambda}$, we generate multiple realizations of $\Omega_{\mathrm{GW}}$ and compute their mean, which serves as an estimator for the expectation value $\overline{\Omega}_{\mathrm{GW}}$. Our MLP is then trained to interpolate this ensemble average, rather than modeling individual realizations. %
To complement this, we also model the intrinsic variance—i.e., the fluctuations of a specific $\Omega_{\mathrm{GW}}$ realisation around $\overline{\Omega}_{\mathrm{GW}}$—as a separate quantity (as described below, one does not need the complexity of a MPL to model the variance; a much simpler interpolation scheme is sufficient). %
The SGWB variance depends on the number of sources contributing to the signal, and therefore on the observation time, with shorter durations yielding larger fluctuations. %
As a result, our approach inherently captures the average behavior across finite-population realizations, which is especially important when the number of sources is small, or equivalently for short observation times, where stochastic fluctuations are more significant.

We compare the average time required to compute the spectrum for a single set of hyperparameters using different methods, as it constitutes the dominant cost in the likelihood evaluation process. Given a population of $10^5$ sources, traditional Monte Carlo approaches typically require on the order of $10^2$ seconds, while reweighting-based methods reduce this significantly, with each computation taking approximately $10^{-1}$ seconds on average. In contrast, our MLP-based approach enables near-instantaneous spectra evaluations, with each taking approximately $10^{-4}$ seconds. All timings were obtained on the same MacBook Pro with an M4 chip, without parallelization or multithreading. While the MLP requires an initial training phase of $\sim$1 hour on the same machine, the cost is quickly amortized in scenarios involving tens or hundreds of thousands of likelihood evaluations.

\section{Population models}
\label{sec:population}

We now introduce the population models we use to describe the redshift and mass distributions of binary black holes. These models will be used to compute the SGWB training sets. %
In this paper, we assume non-spinning\footnote{We have independently verified that assuming the black-hole spin distribution most recently inferred by the LVK collaboration~\cite{2021ApJ...913L...7A} does not appreciably change GWB estimates.} %
black holes in non-eccentric binaries, hence mass and redshift are the key parameters that capture the complexity of the population. %
A summary of the population model parameters $\mathbf{\Lambda}$ used in this work is provided in Table~\ref{pop_params}; some representative models are illustrated in Fig.~\ref{fig:population_models}. 
These models for the SGWB are further illustrated in \ref{appendix:variations}, where we vary individual population parameters while fixing others at reference values. %

\begin{table}
\caption{\label{pop_params}Summary of population model parameters. The table includes the ``Madau-Dickinson'' (MD) redshift model, the ``power-law plus peak'' (PP) mass model, and the ``broken power-law'' (BPL) mass model.}
\vspace{0.3cm}
\small
\centering
\begin{tabular}{@{}ll}
\br
\multicolumn{2}{@{}l}{\bf Redshift model: Madau-Dickinson} \\
\mr
$\gamma$ & Power-law index at low redshift \\
$\kappa$ & Power-law index at high redshift \\
$z_{\mathrm{peak}}$ & Redshift at which the merger rate peaks \\
$\mathcal{R}_0$ & Local merger rate \\
\br
\multicolumn{2}{@{}l}{\bf Mass model: power-law plus peak} \\
\mr
$\alpha$ & Power-law index of the primary-mass distribution \\
$\beta$ & Power-law index of the mass-ratio distribution \\
$m_{\mathrm{min}}$ & Minimum-mass cutoff of the primary-mass distribution \\
$m_{\mathrm{max}}$ & Maximum-mass cutoff of the primary-mass distribution \\
$\lambda_{\mathrm{peak}}$ & Fraction of the population in the Gaussian component \\
$\mu_{\mathrm{peak}}$ & Mean of the Gaussian component in the primary-mass distribution \\
$\sigma_{\mathrm{peak}}$ & Width of the Gaussian component in the primary-mass distribution \\
$\delta_m$ & Width of mass tapering at the low-mass end \\
\br
\multicolumn{2}{@{}l}{\bf Mass model: broken power law} \\
\mr
$\alpha_1$ & Power-law index of the primary mass below $m_{\mathrm{break}}$ \\
$\alpha_2$ & Power-law index of the primary mass above $m_{\mathrm{break}}$ \\
$\beta$ & Power-law index of the mass-ratio distribution \\
$m_{\mathrm{min}}$ & Minimum mass of the primary-mass distribution \\
$m_{\mathrm{max}}$ & Maximum mass of the primary-mass distribution \\
$b$ & Fractional position of the break in the allowed mass range \\
$\delta_m$ & Width of mass tapering at the low-mass end \\
\br
\end{tabular}
\end{table}

\subsection{Redshift model: Madau-Dickinson} The merger rate $\mathcal{R}$ of binary black holes per unit comoving volume $V_c$ and unit source frame time is modeled after the star-formation rate density as described by Madau and Dickinson (MD)~\cite{2014ARA&A..52..415M}:
\begin{equation}
    {\cal R}(z|{\cal R}_0, \gamma, \kappa, z_{\mathrm{peak}}) = {\cal R}_0  (1+z)^{\gamma}
    \frac{1 + \left(\frac{1}{1+z_{\mathrm{peak}}}\right)^{\kappa}}{1 + \left(\frac{1+z}{1+z_{\mathrm{peak}}}\right)^{\kappa}}\,.
\end{equation}
which depends on four population parameters: the local merger rate ${\cal R}_0 $, the low-redshift spectral index $\gamma$, the high-redshift spectral index $\kappa$, and the location of the turnover $z_{\rm peak}$. 
The redshift probability distribution is then given by \cite{2018ApJ...863L..41F, Renzini:2024hiu}:
\begin{equation}
    p(z) \propto \frac{1}{1+z} \frac{dV_c}{dz} {\cal R}(z|{\cal R}_0, \gamma, \kappa, z_{\mathrm{peak}}),
\end{equation}
We assume Planck 18 cosmology~\cite{Planck:2018vyg} for all calculations.

\begin{figure}
    \centering
    \includegraphics[width=\columnwidth]{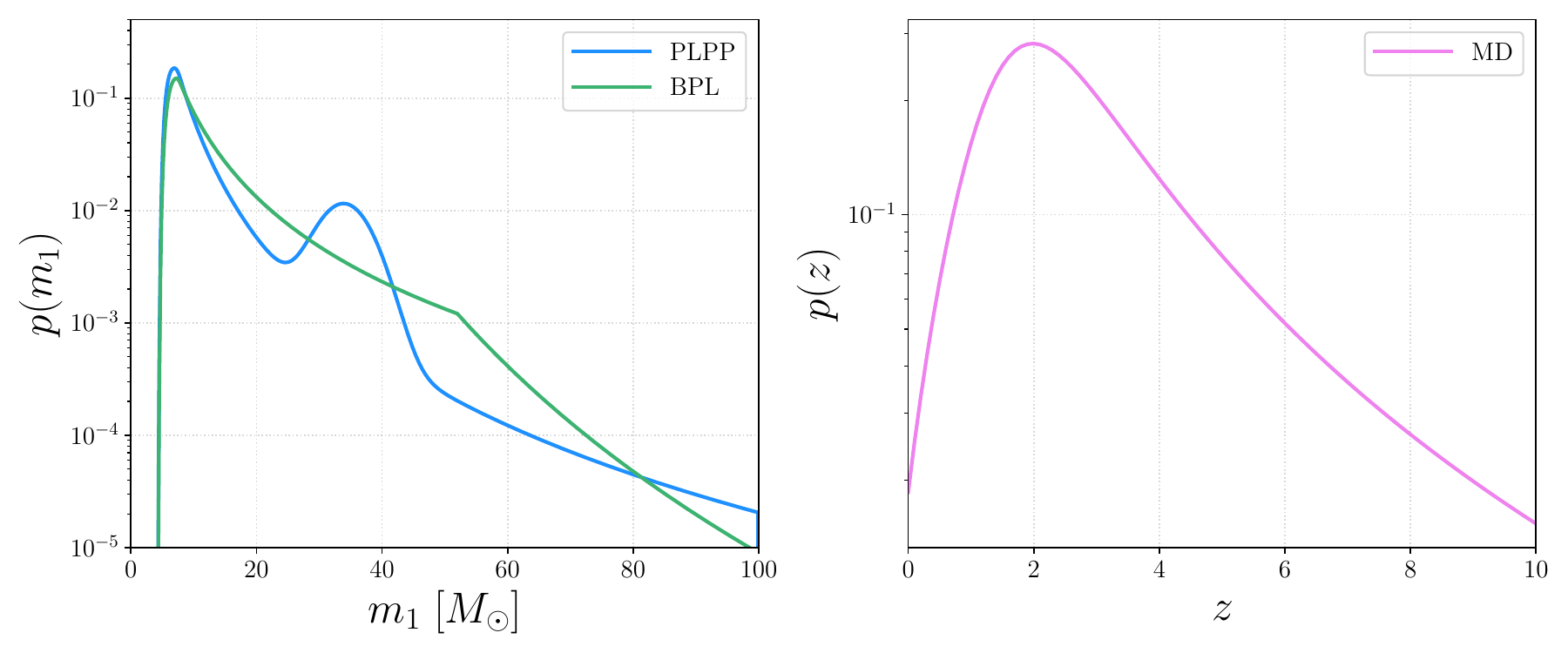}
    \caption{Black-hole binary population models. \textit{Left:} Primary-mass distributions for both the PP (blue) and BPL (green) models. For this example, we set $\alpha = 3.5$, $\delta_m = 4.5$, $\lambda_{\mathrm{peak}} = 0.04$, $m_{\mathrm{max}} = 100$, $m_{\mathrm{min}} = 4$, $m_{\mathrm{peak}} = 34$, $\sigma_{\mathrm{peak}} = 4$ for PP, and  $\alpha_1 = 2.5$, $\alpha_2 = 7.5$, $b = 0.5$, $\delta_m = 4.5$, $m_{\mathrm{max}} = 100$, $m_{\mathrm{min}} = 4$ for BPL. \textit{Right:} Redshift distribution for the MD model (pink) assuming $\gamma = 3.2$, $\kappa = 5.9$, and $z_{\mathrm{peak}} = 1.9$.} 
    \label{fig:population_models}
\end{figure}

\subsection{Mass model: Power-law plus peak} The power-law plus peak (PP) model is a popular \cite{2021ApJ...913L...7A,2023PhRvX..13a1048A} parametrization for the black-hole binary mass spectrum. The primary-mass distribution is a mixture of a truncated power-law component (${\cal P}$) and a Gaussian component (${\cal G}$)
\begin{align}
        p(m_1 | \lambda_{\mathrm{peak}}, \alpha, m_{\mathrm{min}}, m_{\mathrm{max}}, \delta_m, \mu_{\mathrm{peak}}, \sigma_{\mathrm{peak}}) = [(1 - \lambda_{\mathrm{peak}}){{\cal P}}(m_1 | -\alpha, m_{\mathrm{max}}) \notag\\
        + \lambda_{\mathrm{peak}}{\cal G}(m_1| \mu_{\mathrm{peak}}, \sigma_{\mathrm{peak}})]S(m_1|m_{\mathrm{min}}, \delta_m)
\end{align}
where the spectral index of the power-law component is $-\alpha$, the high-mass cutoff of the power-law component is $m_{\rm max}$, the mean of the Gaussian component is $\mu_{\mathrm{peak}}$, the standard deviation of the Gaussian component is $\sigma_{\mathrm{peak}}$, and the mixing fraction is $\lambda_{\mathrm{peak}}$. The smoothing term
\begin{align} \label{smoothing_function_PLPP}
    S(m \mid m_{\mathrm{min}}, \delta_m) &= 
    \begin{cases}
        0 & {\rm if\;\;}m < m_{\mathrm{min}}\,, \\
        \left[f(m - m_{\mathrm{min}}, \delta_m) + 1\right]^{-1} & {\rm if\;\;}m_{\mathrm{min}} \leq m < m_{\mathrm{min}} + \delta_m\,, \\
        1 & {\rm if\;\;}m \geq m_{\mathrm{min}} + \delta_m\,,
    \end{cases}
\\
    f(m', \delta_m) &= \exp\left(\frac{\delta_m}{m'} + \frac{\delta_m}{m' - \delta_m}\right)\,,
\end{align}
regularizes the low-mass end of the distribution.
The distribution of the mass ratio $q=m_2/m_1\leq 1$ conditioned on that of the primary mass is modeled by a smoothed power law
\begin{equation} \label{ratio_dist_PLPP}
    p(q \mid m_1, \beta, m_{\mathrm{min}}, \delta_m) \propto q^{\beta}\, S(q m_1 \mid m_{\mathrm{min}}, \delta_m)\,.
\end{equation}

\subsection{Mass model: broken power-law} For comparison, we also consider an alternative model for the mass distribution: a broken power law (BPL). However, we note that Ref.~\cite{2021ApJ...913L...7A} found this model to be disfavored relative to PP. In this prescription, one has
\begin{align}
    p(m_1 \mid \alpha_1, \alpha_2, m_{\mathrm{min}}, m_{\mathrm{max}}) &= 
    \begin{cases}
        m_1^{-\alpha_1}\, S(m_1 \mid m_{\mathrm{min}}, \delta_m) & {\rm if\;\;}m_{\mathrm{min}} < m_1 \leq m_{\mathrm{break}}\,,\\
        m_1^{-\alpha_2}\, S(m_1 \mid m_{\mathrm{min}}, \delta_m) &  {\rm if\;\;}m_{\mathrm{break}} < m_1 \leq m_{\mathrm{max}}\,, \\
        0 & \text{otherwise}\,,
    \end{cases}
	\\
    m_{\mathrm{break}} &= m_{\mathrm{min}} + b (m_{\mathrm{max}} - m_{\mathrm{min}})\,,
\end{align}
where $\alpha_1$ and $\alpha_2$ are the spectral indexes for low and high masses, respectively, and $b$ sets the position of the transition point within the allowed mass range. The smoothing function and the mass ratio distribution are the same as in Eq.~(\ref{smoothing_function_PLPP}) and Eq.~(\ref{ratio_dist_PLPP}), respectively. 

\section{Interpolation}
\label{sec:mlp}

\subsection{SGWB ensemble average}

We model the expected SGWB spectrum using MLPs implemented in \textsc{PyTorch}~\cite{2019arXiv191201703P}. Each MLP is trained to predict the ensemble-averaged power spectrum $\overline{\Omega}_{\mathrm{GW}}(\mathbf{\Lambda}; f)$ at a given frequency $f$ as a function of the population hyperparameters $\mathbf{\Lambda}$. 

\begin{figure}
    \centering
    \includegraphics[width=\linewidth]{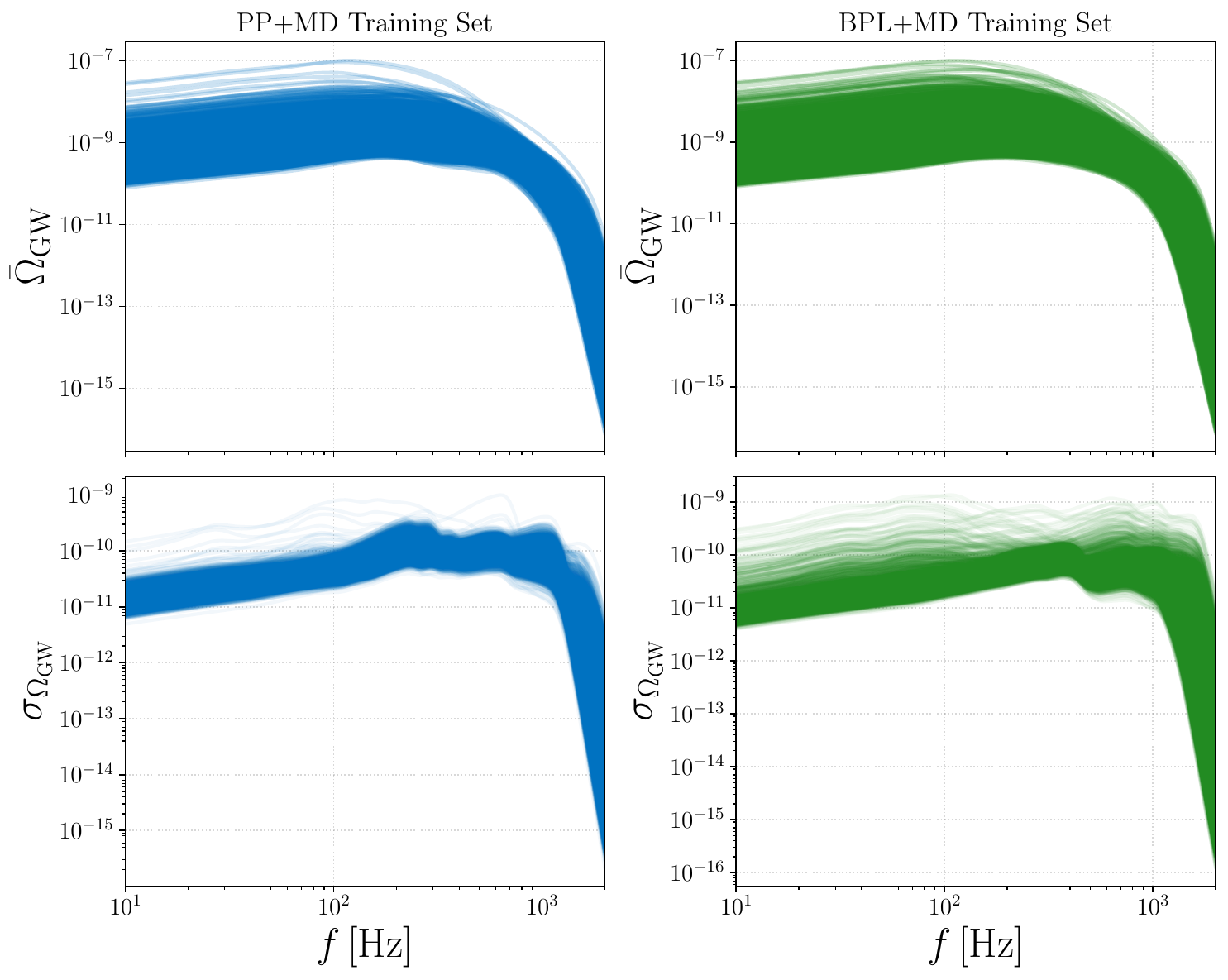}
    \caption{Mean (top panels) and variance (bottom panels) of the SGWB power spectra $\Omega_{\mathrm{GW}}(f)$ computed over 100 realizations for each hyperparameter configuration in the training set only. Results are shown separately for the PP+MD (left panels, blue) and BPL+MD (right panels, blue) models. 
    The spectra span 400 logarithmically spaced frequencies between 10~Hz and 2000~Hz, capturing the sensitivity band of ground-based 3G detectors.
    }
    \label{fig:spectra_training_set}
\end{figure}

To generate the training set, we begin by sampling 8000 distinct combinations of the hyperparameters $\mathbf{\Lambda}$ for each population model. Samples for the PP mass model are drawn from the posteriors available in the LVK data release at Ref.~\cite{gwtc3_zenodo}, while samples for the BPL mass model are available as a part of the example sample sets in the {\sc popstock} package repository~\cite{popstock_repo}. As for the MD model, we use ${\cal R}_0$ and $\gamma$ samples from  Ref.~\cite{gwtc3_zenodo}, while for $\kappa$ and $z_{\mathrm{peak}}$, which are not captured in the simpler redshift model used in Ref.~\cite{2023PhRvX..13a1048A}, we assume Gaussian priors centered at 5.6 and 1.9, respectively, with standard deviations of 1.0 and 0.3 respectively.
        \paragraph{Spectral generation}
        To compute the SGWB spectra of a specific set of hyperparameters, we begin by drawing 100 independent sets of samples from a common fiducial binary population distribution. This fiducial distribution is shared across all hyperparameter configurations within a given model. For each of the 8000 sampled hyperparameter combinations, we compute 100 independent realizations of the SGWB spectrum by reweighting these 100 fiducial sets of samples in a one-to-one manner: the $i$-th realization of a given hyperparameter set is obtained by reweighting the $i$-th fiducial set of samples.
        The fiducial distribution used to generate the base samples is defined, for the PP+MD model, by the parameters: $\alpha = 2.5$, $\beta = 1$, $\delta_m = 3$, $\lambda_{\rm peak} = 0.04$, $m_{\rm min} = 4$, $m_{\rm max} = 100$, $\mu_{\rm peak} = 33$, $\sigma_{\rm peak} = 5$, $\gamma = 2.7$, $\kappa = 5.6$, $z_{\rm peak} = 1.9$, and $\mathcal{R}_0 = 15$. An analogous fiducial distribution is defined for the BPL+MD model, with the same common parameters as well as $\alpha_1 = 2$, $\alpha_2 = 1.4$, $\delta_m = 4.5$, $b = 0.4$, and $\mathcal{R}_0 = 16$. On average, the effective number of binaries used in the reweighting process is $N_{\rm eff} \simeq 2 \times 10^4$. Spectra are computed at 400 logarithmically spaced frequencies between 10 Hz and 2000 Hz, using the waveform approximant \textsc{IMRPhenomD}\footnote{Although this approximant is somewhat outdated and neglects spin precession and eccentricity, its use remains prevalent in many studies of the astrophysical stochastic gravitational-wave background \cite{isotropic_O3}. This is largely due to its computational efficiency and the fact that previous comparisons, such as those in \cite{2024A&A...691A.238R}, have indicated that the overall shape and amplitude of the predicted stochastic background spectrum are not dramatically altered by employing more complex waveform approximants. See also \ref{appendix:waveform_model_sensitivity_check} for a comparison between spectra calculated using \textsc{IMRPhenomD} and \textsc{IMRPhenomXAS}. Therefore, for the scope of this analysis, \textsc{IMRPhenomD} is deemed an adequate choice.}.
        \paragraph{Computation of mean and variance}
        For each of the 8000 hyperparameter combinations, we compute the mean $\overline{\Omega}_{\mathrm{GW}}$ and standard deviation $\sigma_{\Omega_{\mathrm{GW}}}$ of the power spectrum across the 100 realizations, for each of the 400 frequencies. The result is thus a dataset of 8000 mean spectra (each with 400 frequency points) and their corresponding variances, for each model. 
    \vspace{0.5\baselineskip}\\
    The full dataset is split as follows: 80\% for training, 20\% for testing. Within the training set, an additional 80/20 split is applied to create training and validation subsets. The mean and variance of the spectra included in the training set are shown in Fig.~\ref{fig:spectra_training_set}.    

Each MLP is trained to predict the ensemble-averaged power spectrum $\overline{\Omega}_{\mathrm{GW}}(\Lambda; f)$ as a function of the population hyperparameters $\Lambda$ and the frequency $f$. The input to the network consists of the frequency $f$ and the hyperparameters describing the source population: 14 total parameters for the PP+MD model and 13 for the BPL+MD model. The output is a single scalar value $\overline{\Omega}_{\mathrm{GW}}$ at that frequency.

Inputs and outputs are preprocessed in log space to have zero mean and unit variance. We train the networks to minimize the mean absolute error (MAE) using the Adam optimizer~\cite{kingma2017} with an initial learning rate of $10^{-3}$, a batch size of 128, and early stopping based on validation loss with a patience of 20 epochs. Training runs for a maximum of 1000 epochs. 

\begin{table}
\caption{\label{mlp_training_results} Test set performance metrics for the PP+MD and BPL+MD models. The table reports the mean absolute error (MAE), coefficient of determination ($R^2$), and Spearman correlation coefficient.}
\vspace{0.3cm}
\small
\centering
\begin{tabular}{@{}ll}
\br
\multicolumn{2}{@{}l}{\bf Model: PP+MD} \\
\mr
MAE & $1.50 \times 10^{-11}$ \\
$R^2$ & $0.974$ \\
Spearman correlation & $0.99996$ \\
\br
\multicolumn{2}{@{}l}{\bf Model: BPL+MD} \\
\mr
MAE & $7.68 \times 10^{-12}$ \\
$R^2$ & $0.994$ \\
Spearman correlation & $0.99996$ \\
\br
\end{tabular}
\end{table}

\begin{figure}
    \centering
    \includegraphics[width=0.8\linewidth]{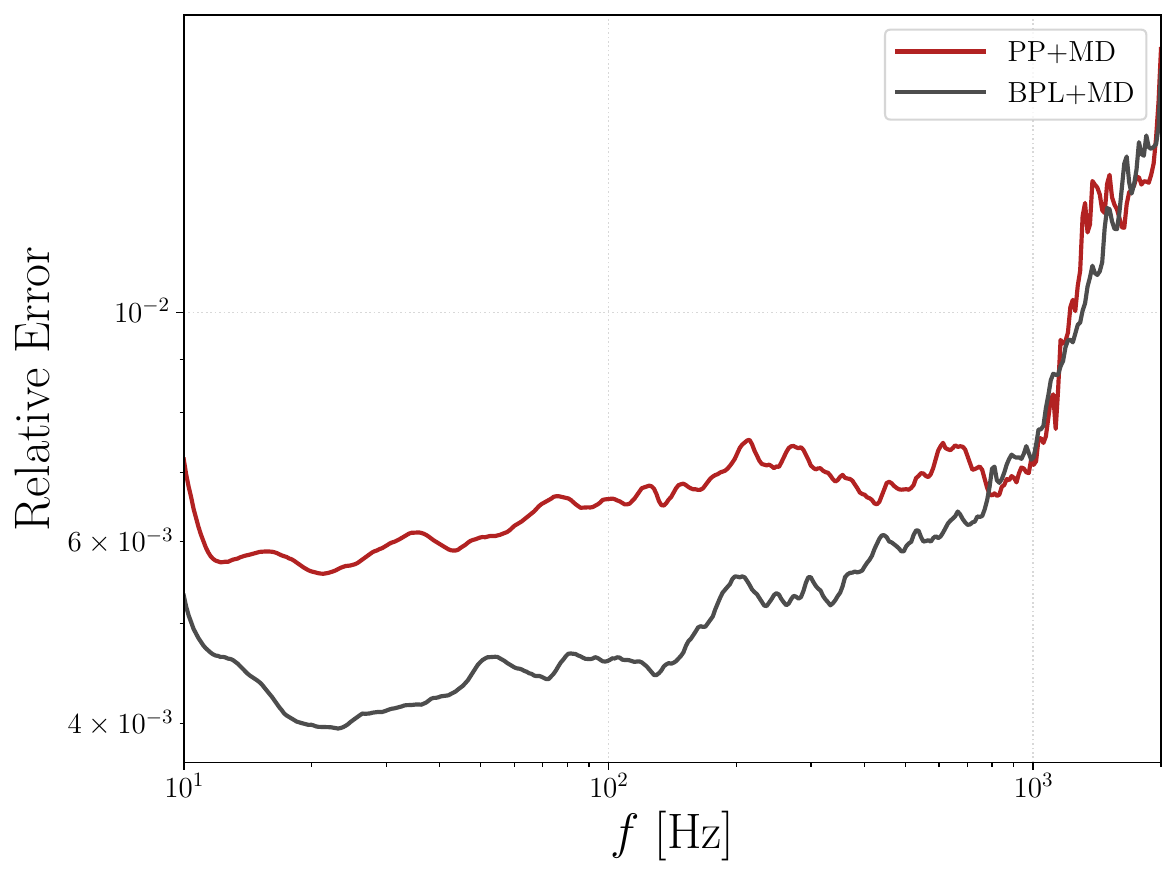}
    \caption{
        Mean relative error of the MLP predictions on the test set as a function of frequency.
        Relative error as a function of frequency for the MLP-based models. The PP+MD model ({red}) and the BPL+MD model ({grey}) are compared across the frequency range of interest. Both axes are on a logarithmic scale. In the frequency range relevant to inference, the relative error remains well below 1\%. 
    }
    \label{fig:relative_error_plot}
\end{figure}

Table~\ref{mlp_training_results} reports the test set performance, including the MAE and the determination coefficient $R^2$. Fig.~\ref{fig:relative_error_plot} shows the mean relative error of the MLP predictions on the test set across frequencies. The error is lowest at the lower end of the frequency range probed (20$-$200 Hz), which is advantageous for inference since detector sensitivity peaks in this range.

\begin{figure}
    \centering
    \includegraphics[width=0.75\textwidth]{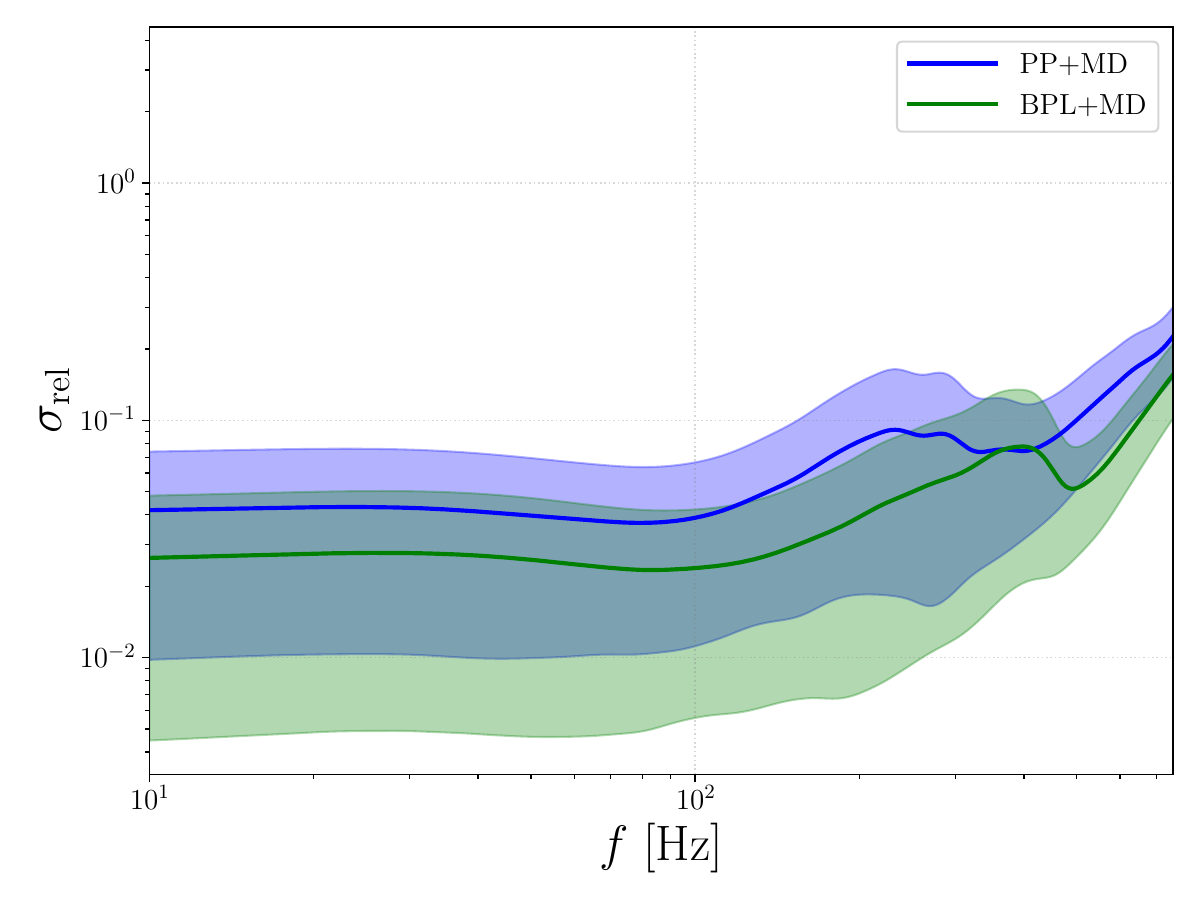}
    \caption{Relative standard deviation $\sigma_{\mathrm{rel}}(f; \mathbf{\Lambda})$, as defined in Eq.~\eqref{eq::rel_std}, computed across 100 realizations for the PP+MD ({blue}) and BPL+MD ({green}) models. Shaded bands show the standard deviation of $\sigma_{\mathrm{rel}}(f; \mathbf{\Lambda})$ across different hyperparameter configurations within each model. The curves are shown up to 750 Hz, which corresponds to the frequency range where the SGWB signal is strongest relative to the expected detector noise (see Sec.~\ref{sec:inference}). The relative variance is lowest at lower frequencies, which is favorable for inference. At higher frequencies, the variance increases as the number of contributing sources drops and the spectrum signal becomes weaker.}
    \label{fig:variance}
\end{figure}

\subsection{SGWB variance}\label{sec:variance}

To complement the mean spectrum prediction, we also quantify the intrinsic variance across the 100 realizations, $\sigma_{\Omega_{\mathrm{GW}}}$. This variance, depicted for the training set in Fig.~\ref{fig:spectra_training_set}, is relevant for inference, as it encodes the stochastic uncertainty associated with a finite number of sources. %
The spectrum variance $\sigma_{\Omega_{\mathrm{GW}}}$ depends on the population parameters $\mathbf{\Lambda}$ %
as well as the number of events, or equivalently the observing time $T_{\rm obs}$, assumed for the calculation of the SGWB. %
One could model the variance with an MPL, as discussed above, conditioned on the number of sources or the observation time; however, we find that simplifying assumptions can be made that obviate the need for this.

We estimate the relative standard deviation %
\begin{equation} \label{eq::rel_std}
    \sigma_{\mathrm{rel}}(f; \mathbf\Lambda) = \frac{\sigma_{\Omega_{\mathrm{GW}}}(f; \mathbf\Lambda)}{\overline{\Omega}_{\mathrm{GW}}(f; \mathbf\Lambda)}\,,
\end{equation} 
across the training set and interpolate its frequency dependence using \textsc{scipy.interp1d}~\cite{2020NatMe..17..261V}. %
The resulting curves for the two mass models are shown in Fig.~\ref{fig:variance}. %
We find that $\sigma_{\mathrm{rel}}(f; \mathbf\Lambda)$ is consistently small, 
lying below 10\% for the majority of the frequency space probed. %
Approximating $\sigma_{\rm rel}(f; \mathbf\Lambda)$ to a constant spectrum leads to a percent-level error on the variance estimate; hence in the inference process we fix it to the mean spectrum 
\begin{equation} \label{eq:rel_std_approx}
    \sigma_{\mathrm{rel}}(f) \simeq \langle\sigma_{\rm rel}(f; \mathbf \Lambda)\rangle_\mathbf\Lambda 
\end{equation}
 which is shown as the solid line is Fig.~\ref{fig:variance}. %
The variance spectrum $\sigma_{\Omega_{\mathrm{GW}}}(f; \mathbf\Lambda)$ is from  Eq.~(\ref{eq::rel_std}) using the MLP prediction for $\overline{\Omega}_{\mathrm{GW}}(f, \mathbf\Lambda)$. %
We verify in the inference process that this approximation is acceptable -- see discussion in Sec.~\ref{sec:inference}.

\section{Inference}
\label{sec:inference}

\subsection{Methodology}

The sensitivity of a gravitational wave detector is characterized by its power spectral density (PSD), which quantifies the noise amplitude as a function of frequency. In this work, we adopt the projected PSDs of ET and CE from Ref.~\cite{2017CQGra..34d4001A}, as implemented in \textsc{PyCBC} \cite{2019PASP..131b4503B}. Specifically, we use the ET-D sensitivity curve, which corresponds to a 10\,km interferometer configuration representing one component of the proposed triangular ET design~\cite{2009CQGra..26h5012F}. For CE, we use the Stage 2 target sensitivity for a 40\,km detector.

We use a Gaussian likelihood to describe the distribution of the SGWB estimator values $\{\hat{\Omega}_{\rm GW}(f_i)\}$ over $N_f$ discrete frequencies $\{f_i\}$. The likelihood is given by~\cite{2012PhRvL.109q1102M}
\begin{equation}
    \mathcal{L}(\hat{\Omega}_{\rm GW}| \Lambda) = \prod_{i=1}^{N_f} \frac{1}{\sqrt{2\pi} \, \sigma_{\mathrm{eff}}(\mathbf\Lambda; f_i)} 
    \exp\left\{ -\frac{1}{2} \frac{\left[\hat{\Omega}_{\rm GW}(f_i) - \overline{\Omega}_{\rm GW}(\mathbf\Lambda; f_i)\right]^2}{\sigma_{\mathrm{eff}}^2(\mathbf\Lambda; f_i)} \right\},
    \label{eq:likelihood}
\end{equation}
where $\overline{\Omega}_{\rm GW}(\mathbf\Lambda; f_i)$ is the MLP prediction for the SGWB energy density spectrum at frequency $f_i$ for a given set of parameters $\mathbf\Lambda$, and $\sigma_{\mathrm{eff}}( \mathbf\Lambda ; f_i)$ is the effective standard deviation combining detector and intrinsic contributions. %
This is defined as: 
\begin{equation}
    \sigma_{\text{eff}}^2(\mathbf \Lambda; f_i) = \sigma_{\rm detector}^2(f_i) + \left[\sigma_{\rm rel}(f_i)\,\overline{\Omega}_{\rm GW}(\mathbf \Lambda; f_i)\right]^2,
    \label{sigmaeff}
\end{equation}
where $\sigma_{\mathrm{detector}}^2(f_i)$, which we define in Eq.~\eqref{eq::sigma-detector}, reflects the uncertainty due to instrumental noise and scales inversely with the observation time, while the second term accounts for the stochastic nature of the SGWB signal, and scales as $1/N_s$, as it is equivalent to the error of a Monte Carlo integral. In a realistic observation, $N_s = \mathcal{R} T_{\mathrm{obs}}$ ($\mathcal{R}$ being the total merger rate integrated over redshift), hence the intrinsic variance of an \emph{observed} signal also scales with $T_{\mathrm{obs}}$. These are assumed to be independent and Gaussian. %

\begin{figure}
    \centering
    \includegraphics[width=0.75\columnwidth]{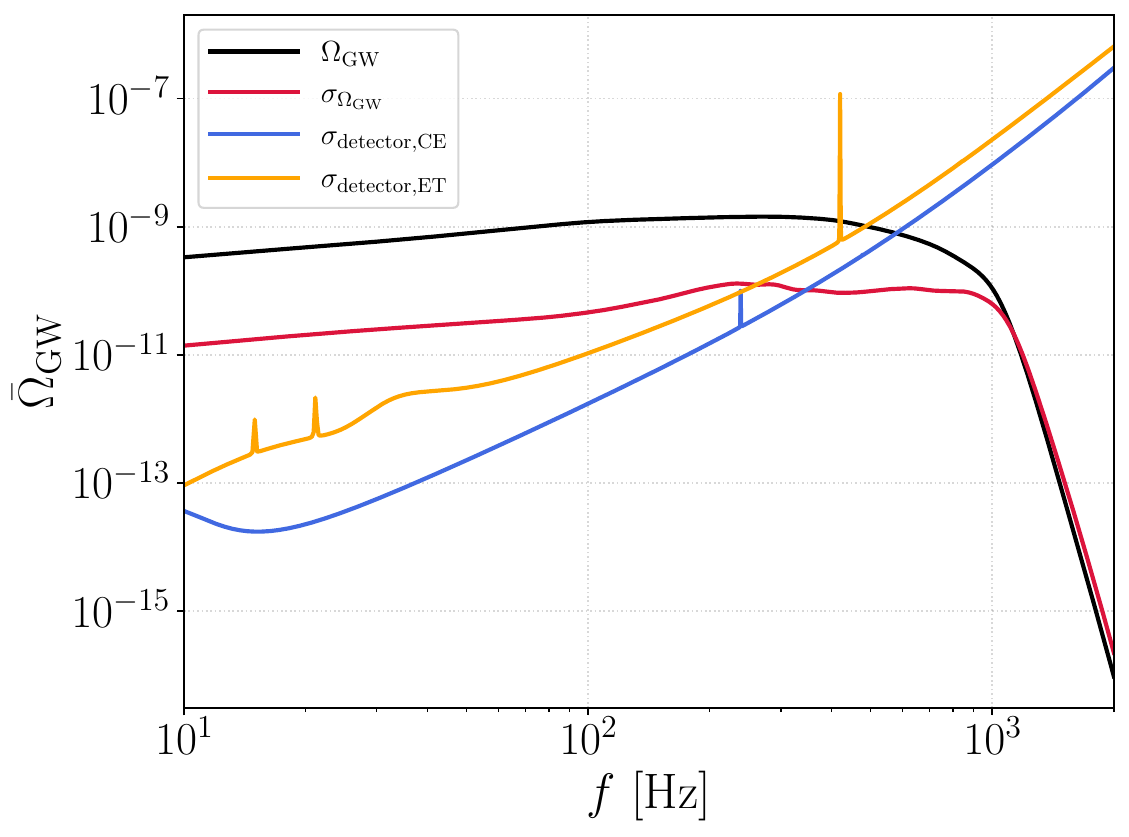}
    \caption{Instrumental uncertainties $\sigma_{\rm detector}$ converted to dimensionless SGWB power spectrum units. Red and blue curves correspond to ET and CE, respectively. The black curve shows an example SGWB realization ${\overline{\Omega}_{\rm GW}}$ from the training set, while the red curve represents its interpolated intrinsic standard deviation $\sigma_{\Omega_{\rm GW}}$. 
    Note the relative amplitude between these quantities: $\sigma_{\Omega_{\rm GW}}$ is orders of magnitude larger than $\sigma_{\rm detector}$ in the lower portion of the frequency spectrum, thus dominating the measurement below $\sim 100$ Hz, while the opposite is true for frequencies higher than $\sim 300$ Hz, for either detector setup.
    }
    \label{fig:psd_sgwb}
\end{figure}

Importantly, the data $\{ \hat{\Omega}_{\rm GW}(f_i) \}$ are the unbiased estimator of the SGWB energy density spectrum. This is constructed from the correlation of strain data over multiple short time segments, which correspond to the weighted average of the squared strain over the total observation time~\cite{Romano:2016dpx}. The associated variance due to the detector noise, in the case of data from a pair of independent colocated, co-aligned correlated with itself, is then~\cite{Romano:2016dpx, Renzini:2023qtj}:  
\begin{equation}\label{eq::sigma-detector}
    \sigma_{\mathrm{detector}}(f) = \frac{10 \pi^2}{\sqrt{2 T_{\mathrm{obs}} \Delta f}} \frac{{\rm PSD}(f)}{3 H_0^2} f^3\,,
\end{equation}
where $H_0$ is the Hubble constant today and we assume a $T_{\mathrm{obs}} = 1\,\mathrm{yr}$. The variances relative to each measurement are shown in Fig.~\ref{fig:psd_sgwb}.
Note that in this simplified approach we assume the detector PSD is perfectly known, that the estimator $\hat\Omega_{\rm GW}$ is unbiased, and that it is possible to correlate colocated, co-aligned detectors without worrying about correlated noise. In a real-data scenario, all of these assumptions will break down, as we discuss in Sec.~\ref{sec:conclusion} and \ref{appendix:noncolocated_detectors}.

We carry out parameter inference using the \textsc{Dynesty} nested sampler~\cite{2020MNRAS.493.3132S}, as implemented in the \textsc{Bilby} library~\cite{2019ApJS..241...27A}, with parallelization enabled. 
The sampler is configured with \texttt{nlive = 1800} and a stopping criterion of \texttt{dlogz = 0.1}, all other parameters are left to default values unless otherwise specified.

\begin{figure}
    \centering
        \includegraphics[width=0.45\textwidth]{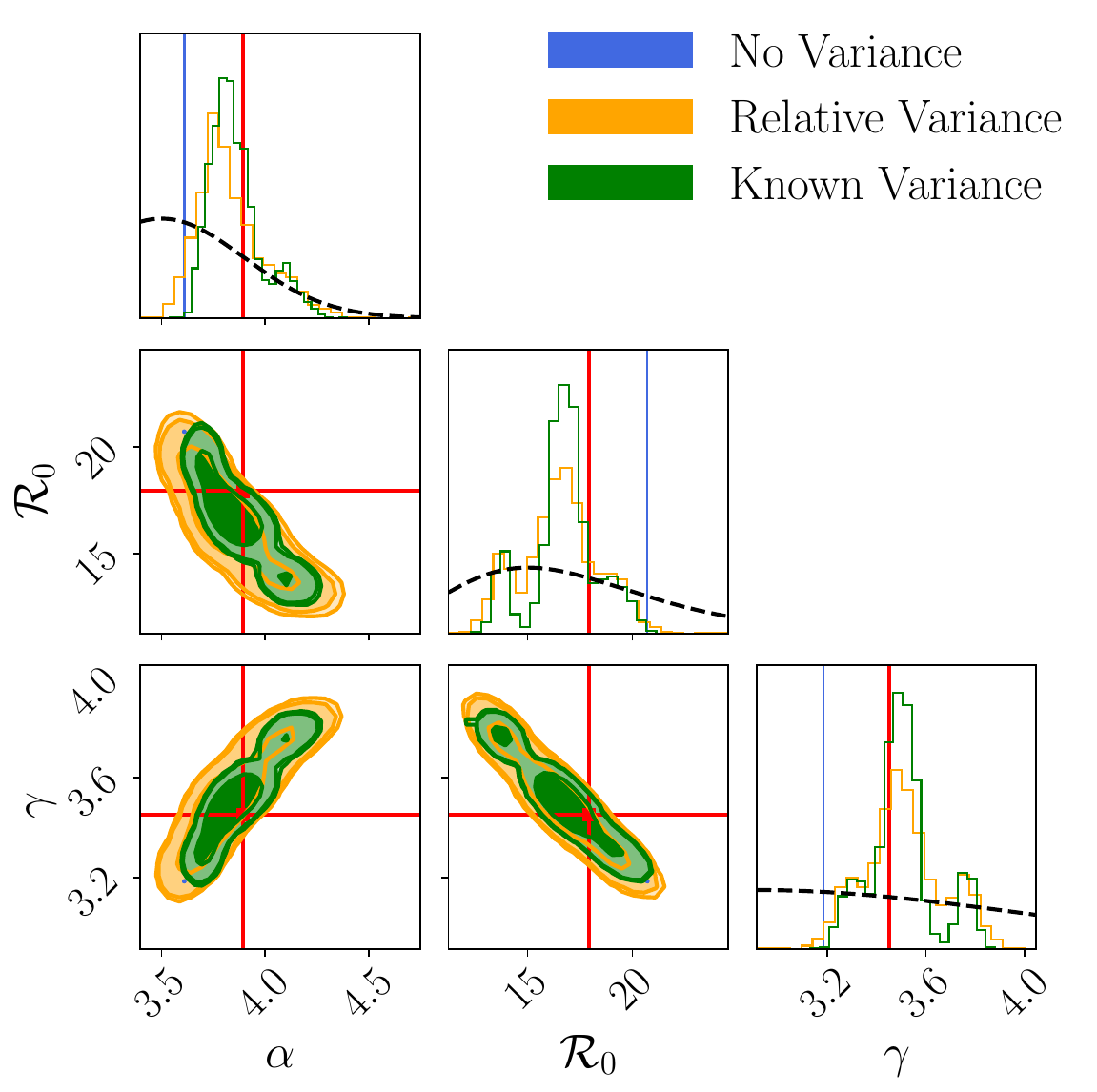}
        \includegraphics[width=0.45\textwidth]{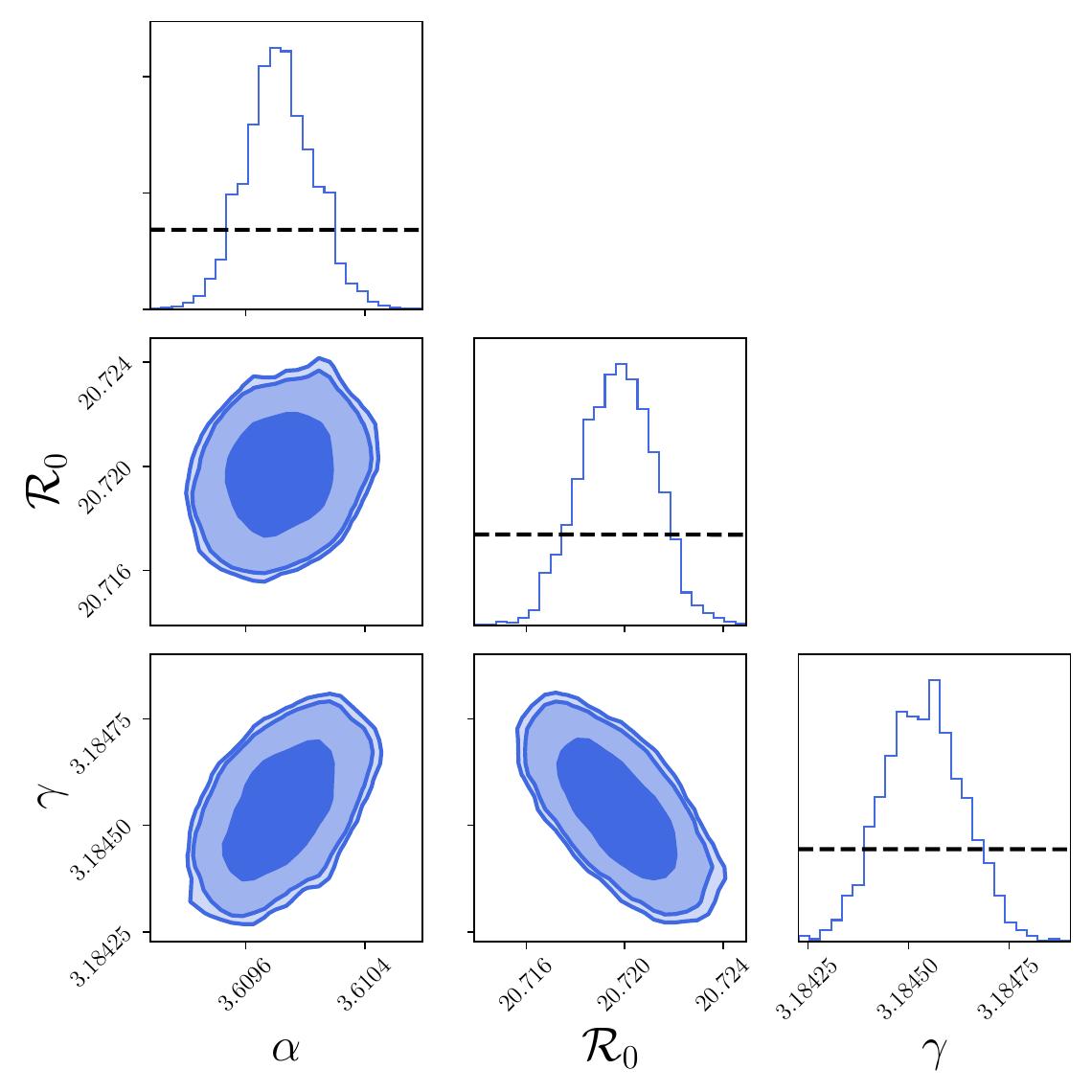}
    \caption{Posterior distributions for a test-set spectrum of the PP+MD model where we fix all parameters except for $\alpha$, $\mathcal{R}_0$, and $\gamma$, inferred under three different treatments of the intrinsic uncertainty. Dashed lines in the one-dimensional marginalized distributions indicate the prior distributions. The left panel compares the full posteriors when intrinsic uncertainty is ignored ({blue}), approximated via interpolation ({orange}), or known for the given spectrum ({green}). The right panel zooms in on the first case, highlighting the biased and overconfident contours resulting from neglecting the intrinsic variance.}
    \label{fig:corner_comparison}
\end{figure}

To illustrate the importance of including the intrinsic variance in the likelihood, we perform parameter inference on a single realization of a spectrum taken from the test set of the PP+MD model, for which both the mean and standard deviation are known and we inject Gaussian noise generated from detector variance. We compare three inference setups: 
\begin{itemize}
    \item [(i)] No variance: the intrinsic uncertainty is entirely neglected, setting $\sigma_{\mathrm{eff}} = \sigma_{\rm detector}$.
    \item [(ii)] Relative variance: the intrinsic uncertainty $\sigma_{\Omega_{\rm GW}}$ is approximated via interpolation of the relative variance, as described in Eq.~\eqref{eq:rel_std_approx}.
    \item [(iii)] Known variance: the known intrinsic variance computed on the test spectrum as a fixed input.
\end{itemize}
The spectrum corresponds to the following source population parameters: $\alpha = 3.89$, $\beta = 2.13$, $\delta_m = 5.52$, $\lambda_{\rm peak} = 0.012$, $m_{\mathrm{max}} = 89.08$, $m_{\rm{min}} = 5.40$, $\mu_{\rm{peak}} = 34.22$, $\sigma_{\mathrm{peak}} = 3.67$, $\mathcal{R}_0 = 17.93$, $\gamma = 3.45$, $\kappa = 4.90$, and $z_{\mathrm{peak}} = 1.69$. In our analysis, all parameters are held fixed except for $\alpha$, $\mathcal{R}_0$, and $\gamma$, which are inferred. As shown in Fig.~\ref{fig:corner_comparison}, ignoring the intrinsic variance leads to significantly biased and overly narrow posterior distributions, which falsely suggest high confidence in incorrect parameter estimates. By contrast, including the intrinsic uncertainty—either approximately through interpolation or exactly using the known variance—results in consistent and accurate posteriors. This demonstrates that even an approximate treatment of intrinsic variance substantially improves inference quality. The effect becomes especially pronounced at shorter observation times, where the detector noise increases and the contribution from intrinsic uncertainty becomes more significant.

\subsection{Inference on the MD Model}
\label{sec:md_article}

We first examine the redshift population parameters in isolation, using the MD model. This provides a clean setup to isolate the effect of redshift evolution on the SGWB signal. The injected signal corresponds to a reference spectrum $\Omega_{\rm GW}$ generated from $10^5$ binary black-hole realizations with population parameters $\alpha = 3.5$, $\beta = 1.0$, $\delta_m = 5.0$, $\lambda_{\rm peak} = 0.03$, $m_{\rm max} = 85.0$, $m_{\rm min} = 5.0$, $m_{\rm pp} = 33.0$, $\sigma_{\rm pp} = 4.5$, $\mathcal{R}_0 = 17$, $\kappa=5.7$, and $z{\rm peak}=1.9$. The signal is injected into Gaussian noise generated from the detector variance, yielding the dataset $\hat{\Omega}_{\mathrm{GW}}$ used for parameter estimation. In the inference, we jointly sample all model parameters using nested sampling. For this analysis, we increase the number of live points to \texttt{nlive = 2500}. To focus on the redshift parameters, which have the strongest impact on the SGWB spectrum, we marginalize over the mass-related parameters by projecting the full posterior distribution onto the redshift subspace. We adopt broader priors for $R_0$, $\gamma$, and $z_{\text{peak}}$ to ensure that the posterior constraints are driven by the data rather than by restrictive prior assumptions, since these parameters are expected to be well constrained by the observed population (see also the discussion in~\ref{appendix:variations}).
\begin{figure}
\centering
\includegraphics[width=\textwidth]{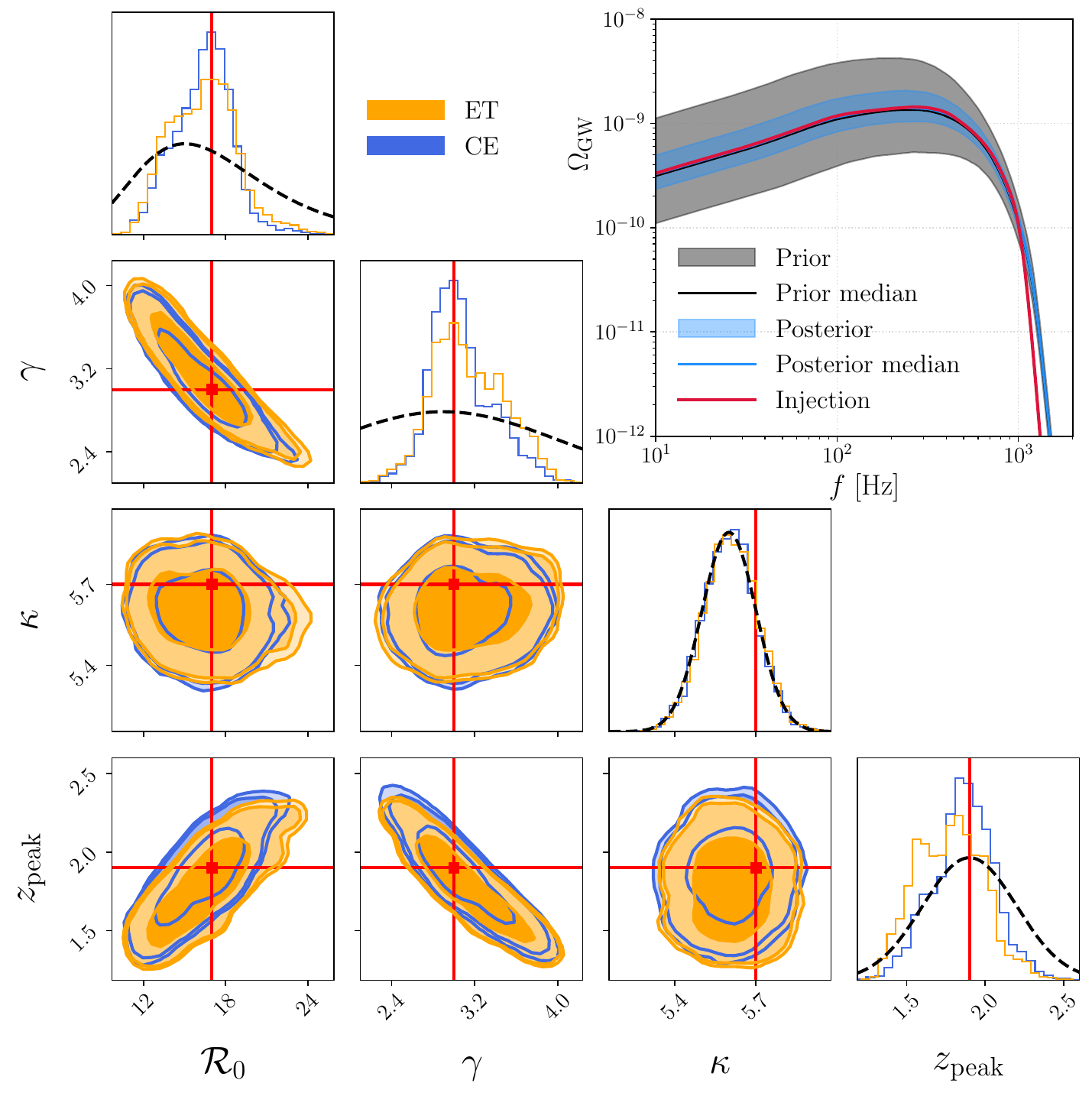}
\caption{
Posterior distributions on the redshift parameters in the MD model, inferred using simulated data from CE ({blue}) and ET ({orange}). The corner plot highlights the consistency between the two detectors in constraining redshift population parameters. Dashed lines in the one-dimensional marginalized distributions indicate the prior distributions using CE simulated data. The inset displays the SGWB spectra predicted by the MLP emulator for CE, showing the median and 90\% credible interval derived from posterior samples ({blue}), as well as from prior samples ({black}) using CE simulated data, while the red curve shows the injected SGWB signal corresponding to a single realization.} 
\label{fig:inference_redshift_md}
\end{figure}

As shown in Fig.\ref{fig:inference_redshift_md}, $\mathcal{R}_0$, $\gamma$, and $z_{\rm peak}$ are the most tightly constrained. This is expected given their dominant impact on the low-frequency end of the SGWB spectrum, where sensitivity is highest. The consistency between CE and ET posteriors confirms the robustness of these constraints. In contrast, $\kappa$ remains poorly constrained due to its limited spectral imprint in the relevant frequency band (see also Fig.~\ref{fig:GWB_variation_plpp_redshift}).

The top right portion of Fig.~\ref{fig:inference_redshift_md} shows an inset comparing the injected and recovered $\Omega_{\rm GW}$ spectrum, and the prior and posterior area. %
As may be observed here, the posterior is significantly constraining and is broadly consistent with the injection. %
Note that at high frequencies ($f > 1000$ Hz), the inference becomes unreliable. Although the model was trained over a broader frequency range, the signal at these high frequencies is strongly dominated by detector noise and lacks sufficient event statistics. As a result, the posterior in this region is prior-dominated and does not provide meaningful constraints.

\subsection{Inference on the PP+MD Model}
\label{sec:plpp_md_article}

\begin{figure}[t]
    \centering
    \includegraphics[width=\textwidth]{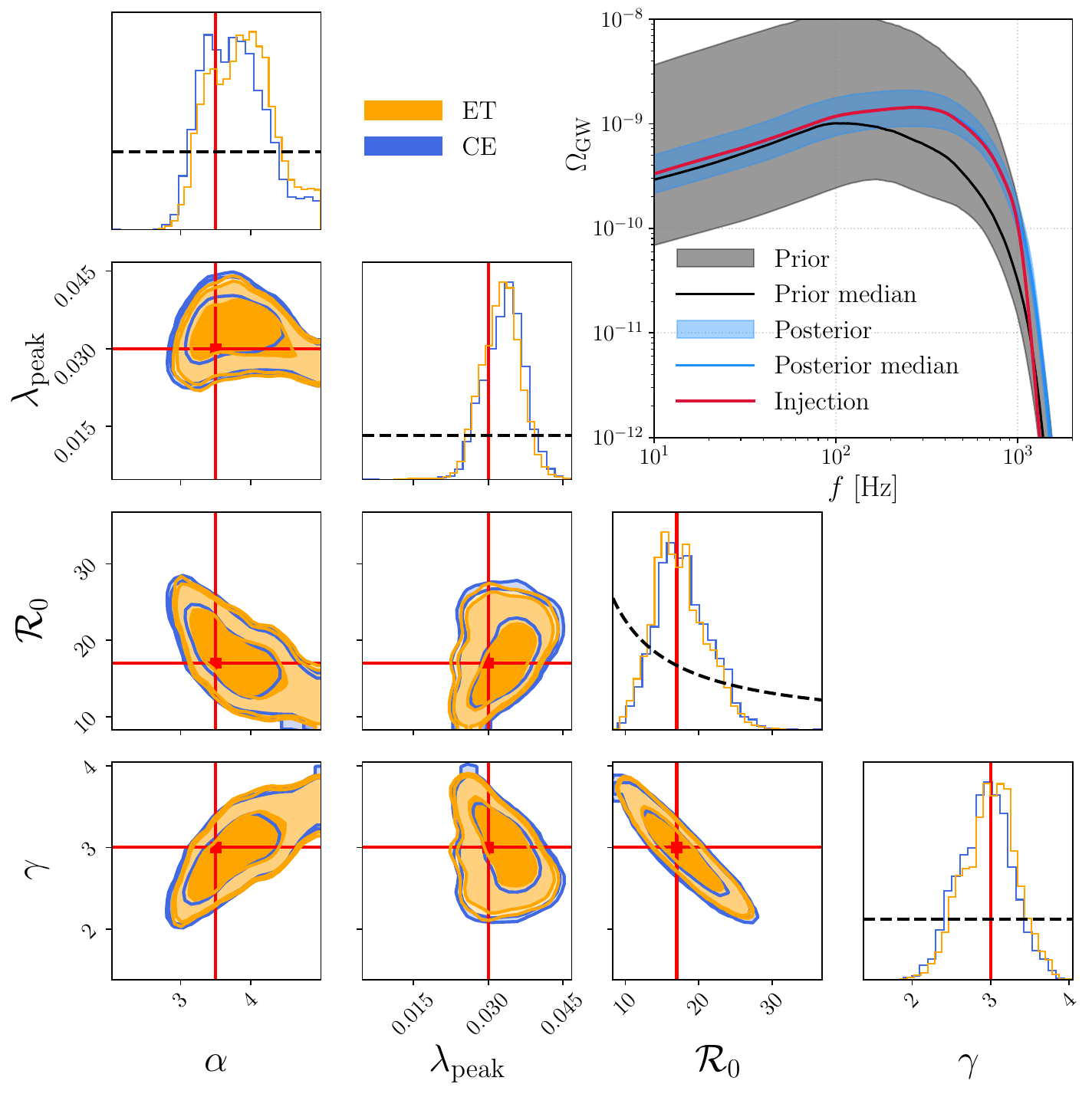}
    \caption{Posterior distributions of $\alpha$, $\lambda_{\rm peak}$, $\mathcal{R}_0$, and $\gamma$ from the PP+MD model inferred using simulated data from CE ({blue}) and ET ({orange}). The corner plot highlights the consistency between the two detectors in constraining the mass and redshift population parameters. Dashed lines in the one-dimensional marginalized distributions indicate the prior distributions. The inset displays the SGWB spectra predicted by the MLP emulator for CE, showing the median and 90\% credible interval derived from posterior samples ({blue}), as well as from prior samples ({black}) using CE simulated data, while the red curve shows the injected SGWB signal corresponding to a single realization.
    }
    \label{fig:inference_plpp}
\end{figure}

We perform parameter inference on the combined PP+MD model using simulated stochastic background data from CE and ET. The injected signal corresponds to the same reference spectrum $\Omega_{\rm GW}$ introduced previously.

\begin{figure}
    \centering
    \includegraphics[width=0.8\columnwidth]{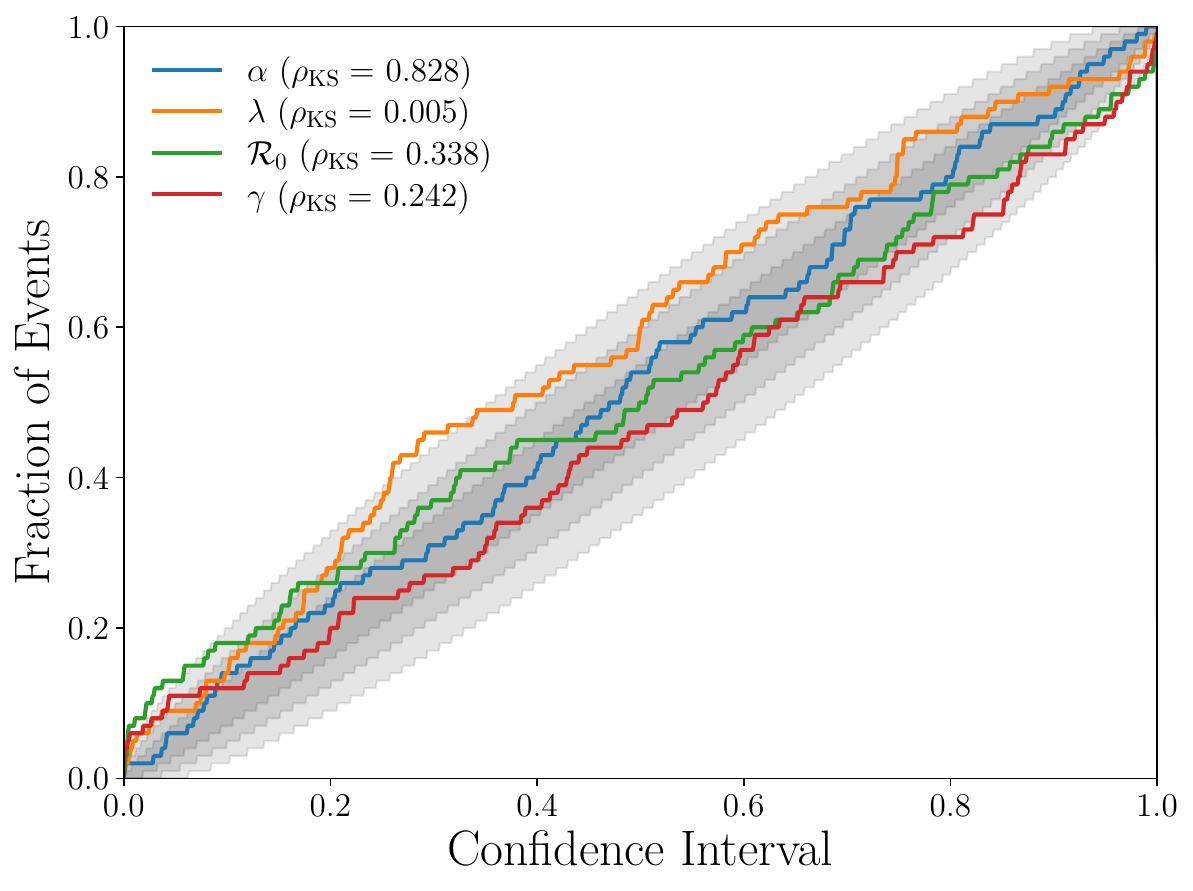}
    \caption{
    Probability-probability test evaluating the calibration of the inferred posteriors for the PP+MD model using simulated CE data. The plot compares the empirical cumulative distribution of true parameter values with the corresponding posterior credible levels, aggregated over 100 simulated injections. A perfectly calibrated inference would produce a curve along the diagonal. The $1\sigma$, $2\sigma$, and $3\sigma$ confidence intervals are indicated by progressively darker shaded regions. Overall, the deviations remain within acceptable bounds, confirming that the inference remains statistically sound for the sample size considered. 
    The numbers in the legend show the $p$-values corresponding to the KS test applied to the individual parameters. 
   }
    \label{fig:pp_plot}
\end{figure}

To avoid performing inference on the entire 12-parameter set for the PP+MD model, as several parameters have a negligible effect on the background, we identify a subset of parameters which effectively dominate the background spectral shape and amplitude: $\mathcal{R}_0$, $\gamma$, $\lambda_{\rm peak}$, and $\alpha$. %
We refer the reader to~\ref{appendix:variations} for a detailed view of the effect of each parameter in the PP+MD model on $\Omega_{\rm GW}$ and to Table \ref{summary_params} for a complete summary indicating which parameters are fixed, marginalized, or constrained in each inference run, considering each parameter individually drawn from the population posteriors published in~\cite{2023PhRvX..13a1048A}. %
In this section, to simplify and speed up the inference, parameters that are not inferred are fixed to their injected values\footnote{We also performed a full inference run including all model parameters, employing priors matching the training priors (LVK population,~\cite{2023PhRvX..13a1048A}). Parameters $\beta$,  $m_{\rm max}$, $\mu_{\rm peak}$, $\sigma_{\rm peak}$ as well as $\kappa$, exhibited posterior distributions indistinguishable from the priors used, indicating they are effectively unconstrained by the data, %
while $m_{\rm min}$ and $\delta_m$ are degenerate, and are discussed in Sec.~\ref{sec:bpl_md_article}.}, while we employ broad, uninformative priors on the selected informative parameters, these were chosen for their demonstrable influence on the spectrum and ability to be constrained by current data,

Results are shown in Figs.~\ref{fig:inference_plpp}.
From these results, we find that $\mathcal{R}_0$ and $\gamma$ are better constrained, consistent with their prominent influence on the low-frequency end of the SGWB spectrum, where the signal is strongest and detector noise is minimal. The ability of SGWB data to constrain the shape of the merger rate history confirms earlier findings such as those in Ref.~\cite{2020ApJ...896L..32C}. The mass distribution parameters $\lambda_{\rm peak}$ and $\alpha$ are also well constrained, reinforcing the sensitivity of the stochastic background signal to the shape of the black-hole mass spectrum, as discussed in Ref.~\cite{2024A&A...691A.238R}.

The top right portion of Fig.~\ref{fig:inference_plpp} shows the inset comparing the injected and recovered $\Omega_{\rm GW}$ spectrum, and the prior and posterior area. %
As in previous cases, the posterior is significantly constraining and is consistent with the injection. %

\begin{figure}
    \centering
    \includegraphics[width=\textwidth]{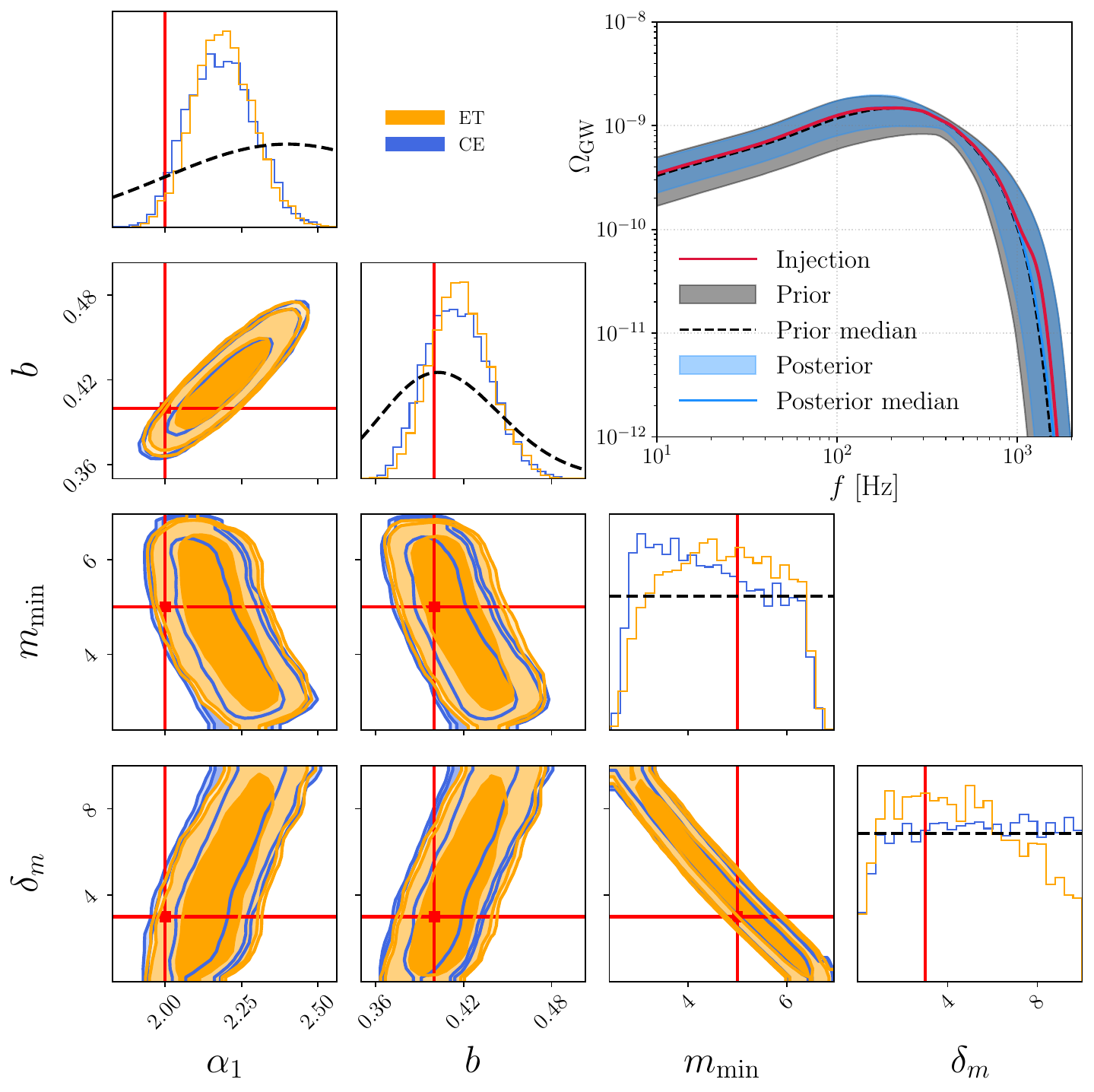}
    \caption{
    Posterior distributions for a subset of BPL+MD model mass parameters from simulated SGWB data from CE ({blue}) and ET ({orange}). Notably, a strong correlation between $m_{\min}$ and $\delta_m$ is evident, reflecting the known degeneracy in shaping the low-mass end of the black hole mass spectrum. The inset displays the SGWB spectra predicted by the MLP emulator for CE, showing the median and 90\% credible interval derived from posterior samples ({blue}), as well as from prior samples ({black}) using CE simulated data, while the red curve shows the injected SGWB signal corresponding to a single realization.} 
    \label{fig:inference_bpl}
\end{figure}

To assess the statistical reliability of our inference pipeline, we perform a posterior predictive test using 100 independent simulations, each generated using parameter values drawn from the prior. For this analysis, the priors are chosen to match the posteriors reported in Ref.~\cite{2023PhRvX..13a1048A}. For each simulated dataset, we perform inference using the CE configuration and evaluate how often the injected (true) parameter values fall within a given posterior credible level. Fig.~\ref{fig:pp_plot} shows the resulting empirical coverage for a subset of representative parameters in the PP+MD model. Despite these small deviations, the overall calibration is statistically acceptable for a sample size of 100 injections, confirming the robustness of our parameter-estimation approach.

\subsection{Inference on the BPL+MD model}
\label{sec:bpl_md_article}

Finally, we present results obtained with the BPL+MD model.
Redshift parameters exhibit constraints similar to those already discussed (see Sec.~\ref{sec:md_article} and Fig.~\ref{fig:inference_redshift_md}); therefore, here we focus on the inference results for the BPL mass distribution parameters and their degeneracies. In this example, we employ as priors the posterior distributions used in the MLP training~\cite{popstock_repo}.

We perform parameter inference for the BPL+MD model using simulated stochastic background data assuming ET and CE sensitivity. The injected signal corresponds to a reference spectrum generated from $10^5$ binary black hole realizations, assuming the hyperparameters: $\alpha_1 = 2.0$, $\alpha_2 = 10.0$, $\beta = 1.0$, break fraction $b = 0.4$, $m_{\min} = 5.0$, $m_{\max} = 90.0$, and $\delta_m = 3.0$, with redshift parameters fixed to $\mathcal{R}_0 = 16.0$, $\gamma = 3.0$, $\kappa = 5.6$, and $z_{\rm peak} = 1.9$.

Fig.~\ref{fig:inference_bpl} presents the posterior distributions for the mass parameters $\alpha_1$, $b$, $m_{\min}$, and $\delta_m$, alongside the corresponding background spectra. The inferred posteriors from both detectors are consistent and reveal the expected degeneracy between $m_{\min}$ and $\delta_m$, c.f. Ref.~\cite{2021ApJ...913L...7A}. This correlation is a well-known characteristic of phenomenological mass distribution models. %
The parameter $\alpha_1$ in the best-inferred mass parameter, similarly to the PP mass model case, while $b$ is also constrained, albeit weakly. %
These appear positively correlated in the posteriors but remain distinguishable.

As with the PP+MD case, parameters that leave only a weak imprint on the power spectrum—particularly at frequencies where the detectors are most sensitive—remain poorly constrained. In particular, mass parameters that primarily influence the high-mass end of the distribution, such as $\alpha_2$, have limited impact within the detectors' optimal frequency band, as illustrated in \ref{appendix:variations}.
Parameters not shown in the posteriors are fixed to their injected values.

The top right portion of Fig.~\ref{fig:inference_bpl} highlights the comparison between the injected and recovered $\Omega_{\rm GW}$ spectrum, and the prior and posterior area. %
We find in this case that the posterior is weakly constrains the prior area, remaining consistent with the injection. %
This is also due to the fact that we employ narrower priors in this analysis, focusing specifically on the mass parameter degeneracies. 

\section{Conclusions}
\label{sec:conclusion}

In this paper, we improve robustness and efficiency of SGWB analysis and offer an efficient alternative to the time-consuming process of population hyper-parameter estimation involving hundreds of thousands of CBC events which is expected in 3G detectors. 
We perform population parameter inference by treating the signal as a loud stochastic background, and infer population parameters from the background spectrum directly assuming an optimal estimator. %
This approach is complementary to population inference involving individual events, and will provide a precious 
control result for compatibility checks and model testing. 

We leverage a series of MLPs trained on LVK’s third observing run (O3) posterior data to efficiently calculate the binary black hole SGWB power spectrum. %
This method allows for quicker likelihood evaluation in Bayesian inference pipelines, significantly accelerating the inference process. %
We have shown that the MLP approach provides reliable estimates of the mean SGWB power spectrum given a set of population hyper-parameters, from which we estimate the associated variance assuming the scaling relation of Eq.~(\ref{eq::rel_std}). %
In this paper, we model the variance dependence on the hyper-parameters by fixing the relative variance (Equation~\ref{eq::rel_std}), hence assuming the dominant term of the dependence is proportional to the SGWB spectrum itself. %

We consider an observing scenario assuming co-aligned, colocated detectors at 3G sensitivities for one year with uncorrelated noise realisations between detectors, and model the SGWB estimator as Gaussian, as commonly done in the literature both in real and mock observing campaigns~\cite{isotropic_O1,isotropic_O2,isotropic_O3, Regimbau:2012ir, Meacher:2015iua, Meacher:2015rex,  Renzini:2023qtj }. %
This implies the likelihood of Eq.~(\ref{eq:likelihood}), where the variance assumed is given by the combination of detector variance and signal variance. This is a key novelty of this approach and is necessary to achieve unbiased inference results in the strong signal scenario. 

We observed that parameters that influence the low-frequency regime, such as the local merger rate $\mathcal{R}_0$ or the slope of the redshift distribution at low redshift $\gamma$, are more easily constrained, whereas parameters affecting higher frequencies are more challenging due to detector noise and weaker spectral features. %
This behaviour aligns with expectations from previous studies~\cite{2024A&A...691A.238R}, which show that the distribution of redshift parameters primarily influences the amplitude of the SGWB spectrum, while the shape is mostly governed by the mass distribution of binary black holes. %
Parameters that are well-constrained are highlighted in Figures~\ref{fig:inference_plpp} and~\ref{fig:inference_bpl}. The two detector set-ups considered here (ET and CE sensitivity) perform extremely similarly, implying that these measurements are limited by the intrinsic variance of the signal. %
On the other hand, we expect higher sensitivities are required to constrain parameters concerning the high-redshift evolution, and parameters which dominate the higher frequencies where the detector noise dominates in this study (Fig.~\ref{fig:psd_sgwb}). %

As previously noted, the assumption of uncorrelated noise for co-located detectors does not readily apply to real data. %
In present ground-based detector analyses, e.g., employing LVK data~\cite{isotropic_O3}, data from non-colocated detectors are cross-correlated to obtain $\hat\Omega_{\rm GW}$ to achieve independent noise. %
This comes with a penalty factor that takes into account the loss of coherence across the light-travel path, i.e. the distance between the detectors. %
This same penalty factor, or {\it overlap factor}, will be relevant for non-colocated 3G detectors such as ET and CE, if data from these were cross-correlated as in~\cite{isotropic_O3}. %
As sensitivity to the isotropic stochastic background is maximised when detectors are co-located, the most efficient characterisation strategy involves detectors that are as close as possible, and independent noise estimation strategies will be transformative for stochastic searches. %
In practice, with 3G detectors the estimator $\hat\Omega_{\rm GW}$ may be constructed in a quite different fashion compared to how it is currently obtained, and may include correlated detector noise mitigation strategies that are still under development -- for example, in the case of triangular detectors, it may be possible to construct null channels~\cite{Muratore:2021uqj} insensitive to GWs which in turn would allow the construction of two uncorrelated colocated detectors. %
Auxiliary noise-monitoring channels will also be fundamental, especially for L-shaped detectors such as CE. %
Furthermore, it is important to note that a source of correlated terrestrial noise will come from magnetic fields~\cite{thraneCorrelatedMagneticNoise2013,himemotoImpactCorrelatedMagnetic2017,coughlinMeasurementSubtractionSchumann2018,janssensImpactSchumannResonances2021,janssensCorrelated11000Hz2023}, particularly relevant for 3G detectors and especially at short distances from each other. Several magnetic noise mitigation strategies have already been discussed in the literature~\cite{thraneCorrelatedNoiseNetworks2014,coughlinMeasurementSubtractionSchumann2018,  himemotoCorrelatedMagneticNoise2019,Meyers:2020qrb}. %
An extension of this work will take into account both the effect of non-colocated detectors and potential correlated noise. %
Due to these considerations, as well as the fact that some parameters are fixed in the inference, this study should not be considered as a quantitative prediction of the capabilities of the ET and CE detectors. %
The value of this work lies in the qualitative assessment of the capabilities of the inference with the MLP model and its ability to correctly reconstruct population parameters in the limit where the SGWB intrinsic noise can dominate. %

Moreover, we recognize that using an MLP interpolator introduces systematic biases, particularly in regions of parameter space with sparse training coverage. While the injection in this work lies well within the training prior, reducing this concern, it remains important to systematically assess emulator-induced bias in future applications. To address this, we plan to adopt simulation-based inference methods, as in \cite{2025ApJ...982...55L}, which provide a principled way to account for model uncertainty and reduce bias.

For future applications, the training dataset can be expanded to include simulated injections that better capture astrophysical uncertainties, for example by using a wider training prior and/or drawing samples from flexible models. This will ultimately improve the model robustness and generalizability. 
Furthermore, a natural next step of this approach is to make the variance a free parameter in the MLP model, allowing for a more flexible and data-driven estimation of the stochastic uncertainty. %
We plan to develop this in an extension of this work.

Although the current MLP efficiently approximates the power spectrum, exploring alternative architectures, such as Bayesian neural networks \cite{2015arXiv150505424B}, normalizing flows \cite{2019arXiv191202762P}, or variational inference techniques \cite{Mould:2025dts}, which can also naturally incorporate detector and signal noise, could provide a more comprehensive approach, by explicitly modeling the variance of the predictions. This would lead to a more reliable uncertainty quantification in the inferred SGWB power spectrum. We leave these extensions for future work.  

\section*{Data availability statement}

All scripts used for generating the datasets, training the models, and performing inference runs are publicly available in the GitHub repository at \url{https://github.com/ggiarda/accelerated-sgwb-pop-inference}.

\section*{Acknowledgments}

We thank Matteo Bonetti for discussions.
G.G., A.I.R, C.P., and D.G. 
are supported by
ERC Starting Grant No.~945155--GWmining, 
Cariplo Foundation Grant No.~2021-0555, 
MUR PRIN Grant No.~2022-Z9X4XS, 
MUR Grant ``Progetto Dipartimenti di Eccellenza 2023-2027'' (BiCoQ),
Italian-French University (UIF/UFI) Grant No.~2025-C3-386,
and the ICSC National Research Centre funded by NextGenerationEU. 
A.I.R. and D.G. are supported by MSCA Fellowship No.~101064542--StochRewind. 
D.G. is supported by MSCA Fellowship No.~101149270--ProtoBH and MUR Young Researchers Grant No. SOE2024-0000125.
Computational work was performed at CINECA with allocations through INFN and Bicocca.

\section*{ORCID iDs}

Giovanni Giarda \orcidlink{0009-0007-3093-7821} \href{https://orcid.org/0009-0007-3093-7821}{https://orcid.org/0009-0007-3093-7821} \\
Arianna Renzini \orcidlink{0000-0002-4589-3987}
 \href{https://orcid.org/0000-0002-4589-3987}{https://orcid.org/0000-0002-4589-3987} \\
Costantino Pacilio \orcidlink{0000-0002-8140-4992} \href{https://orcid.org/0000-0002-8140-4992}{https://orcid.org/0000-0002-8140-4992} \\
Davide Gerosa \orcidlink{0000-0002-0933-3579} \href{https://orcid.org/0000-0002-0933-3579}{https://orcid.org/0000-0002-0933-3579}

\section*{References}
\bibliographystyle{iopart-num}
\bibliography{accstoch}

\appendix
\section{Hyperparameter variation}
\label{appendix:variations}

We computed reference spectra using the parameters in Table \ref{tab:params_variations}. To understand the individual impact of each hyperparameter, we then varied each parameter within the LVK inference run posteriors~\cite{2023PhRvX..13a1048A} for the PP+MD model and within the range of the example sample sets in the {\sc popstock} package repository~\cite{popstock_repo} for the BPL+MD model, while keeping the others fixed at their reference values. %

\begin{figure}
    \centering
    \includegraphics[width=\textwidth]{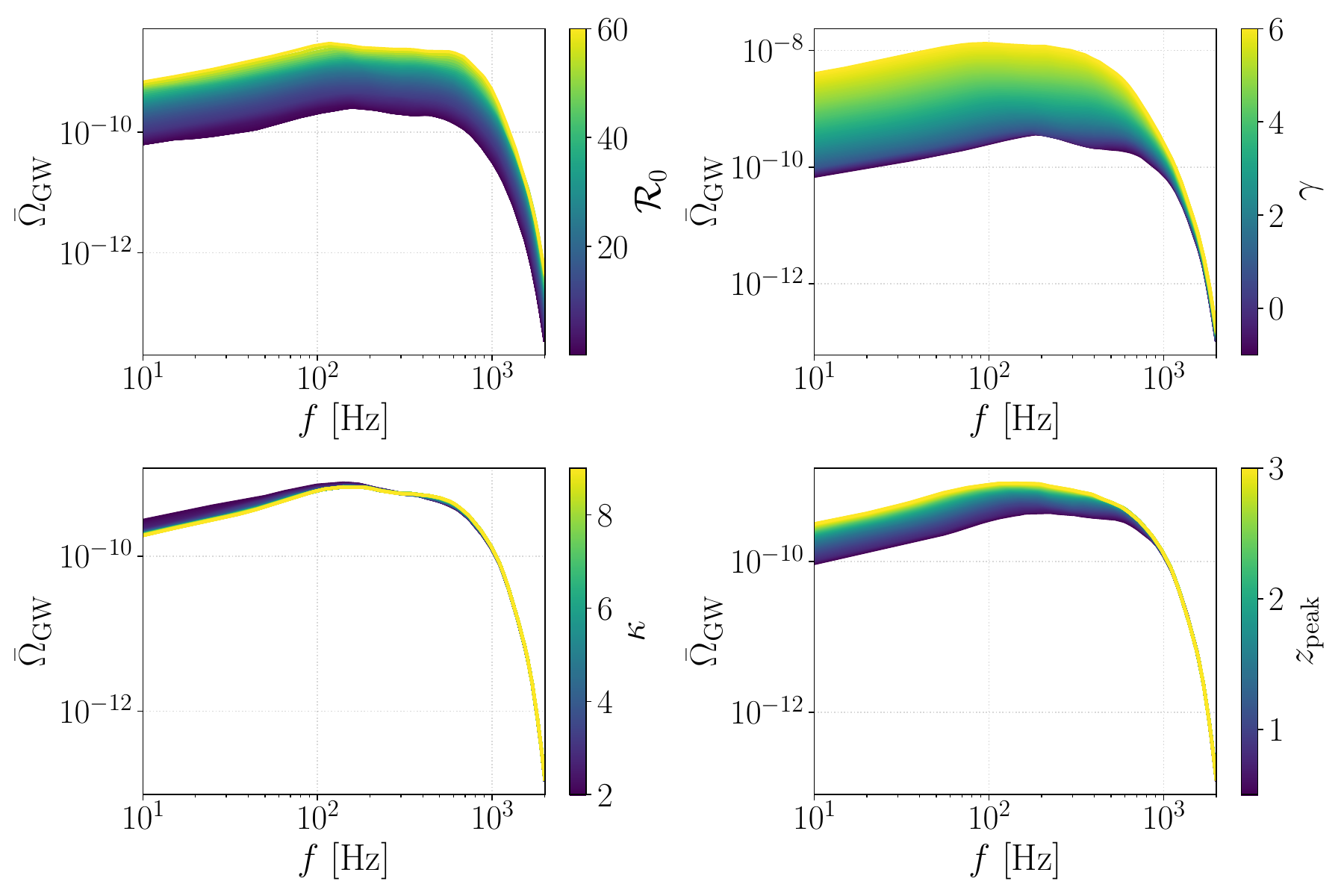}
    \caption{Impact of redshift parameters on the predicted SGWB spectrum for the PP+MD model. The varying parameter ($\mathcal{R}_0$, $\gamma$, $\kappa$, or $z_{\rm peak}$) is indicated next to the color bar, with different color curves indicating different values. All other parameters are kept fixed at their reference values. We observe that redshift parameters, particularly $\mathcal{R}_0$ and $\gamma$ (upper right and left) have a significant impact on the strength of the signal, while $\kappa$ (bottom right) exhibits a comparatively limited influence, which influence the inference process as further discussed in Sec.~\ref{sec:md_article}.
    }
    \label{fig:GWB_variation_plpp_redshift}
\end{figure}

\begin{figure}
    \centering
    \includegraphics[width=0.95\textwidth]{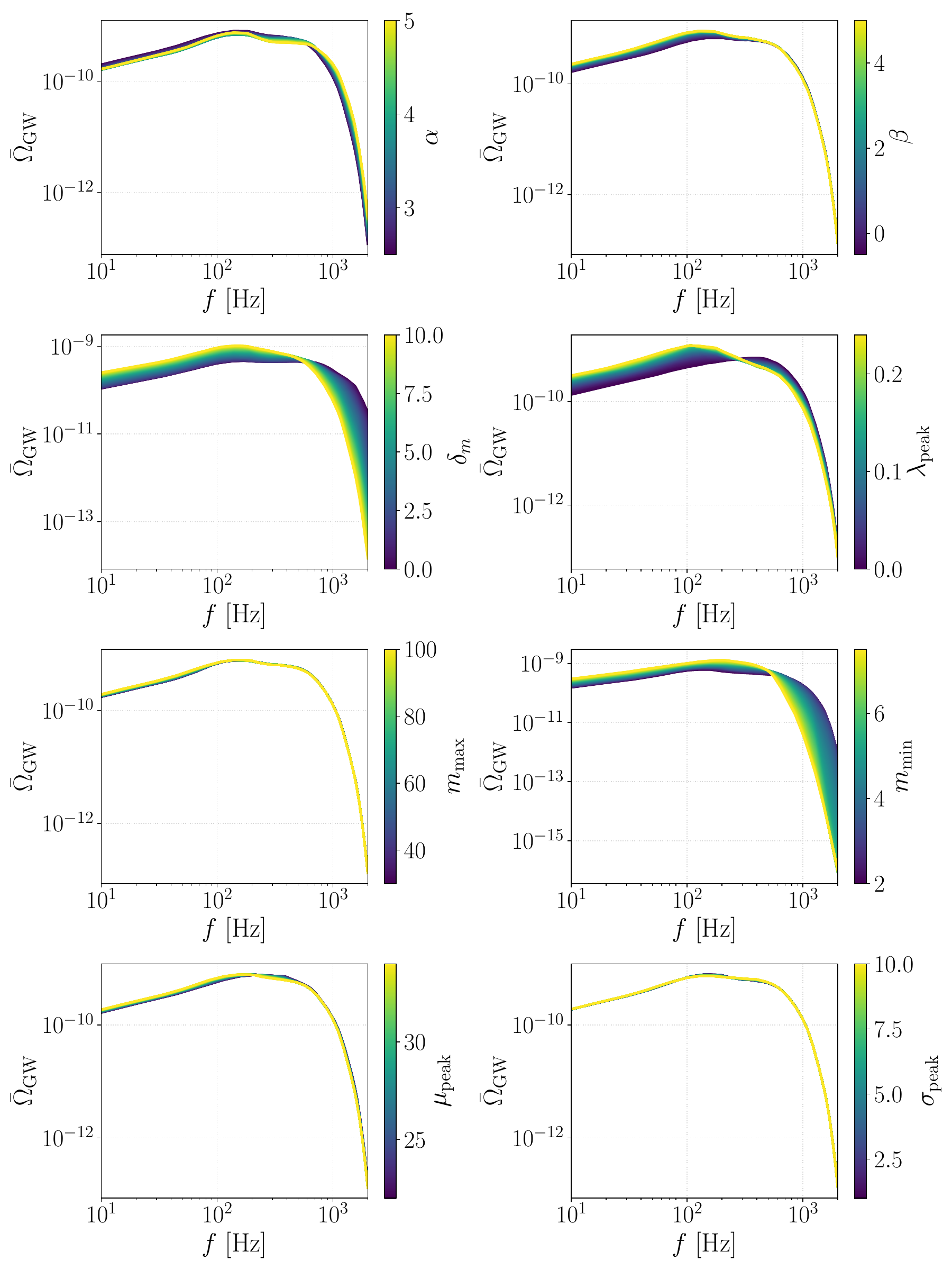}
    \caption{Impact of mass parameters on the predicted SGWB spectrum for the PP+MD model. Each panel illustrates the SGWB spectrum $\bar \Omega_{\text{GW}}(f)$ as a function of frequency $f$. The varying parameter ($\alpha$, $\beta$, $\delta_m$, $\lambda_{\rm peak}$, $m_{\max}$, $m_{\min}$, $\mu_{\rm peak}$, or $\sigma_{\rm peak}$) is indicated next to the color bar in each panel, with different color curves representing different values. All other parameters are kept fixed at their reference values. We observe that parameters like $\mu_{\rm peak}$ and $\sigma_{\rm peak}$ have a relatively low impact on the spectrum, suggesting they will be challenging to constrain directly from the SGWB signal. Additionally, the panels for $m_{\min}$ and $\delta_m$ illustrate the degeneracy between these parameters, as previously discussed in Sec.~\ref{sec:bpl_md_article}. The seemingly limited impact of $\alpha$ is primarily due to the stringent constraints on its variation range, derived from LVK posteriors. While $m_{\max}$ also shows a very limited impact, this isn't problematic for inference as it can often be fixed to fiducial values. 
    }
    \label{fig:GWB_variation_plpp_mass}
\end{figure}

\begin{figure}
    \centering
    \includegraphics[width=\textwidth]{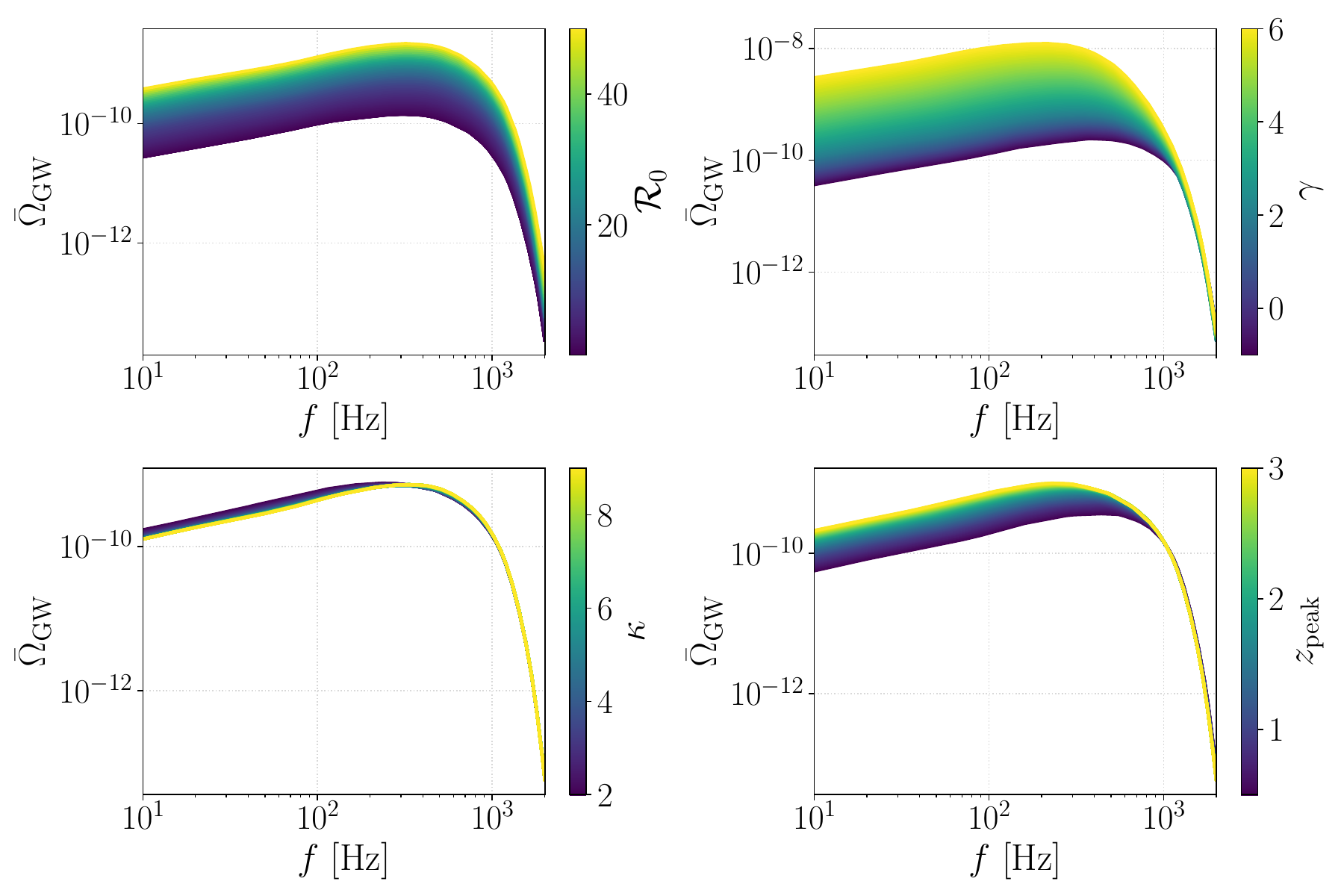}
    \caption{Impact of redshift variation on the predicted SGWB spectrum for the BPL+MD model. The varying parameter ($\mathcal{R}_0$, $\gamma$, $\kappa$, or $z_{\rm peak}$) is indicated next to the color bar, with different color curves indicating different values. All other parameters are kept fixed at their reference values. We notice that the results  are consistent with those shown for the PP+MD model.
    }

    \label{fig:GWB_variation_bpl_redshift}

\end{figure}

\begin{figure}
     \centering
     \includegraphics[width=0.95\textwidth]{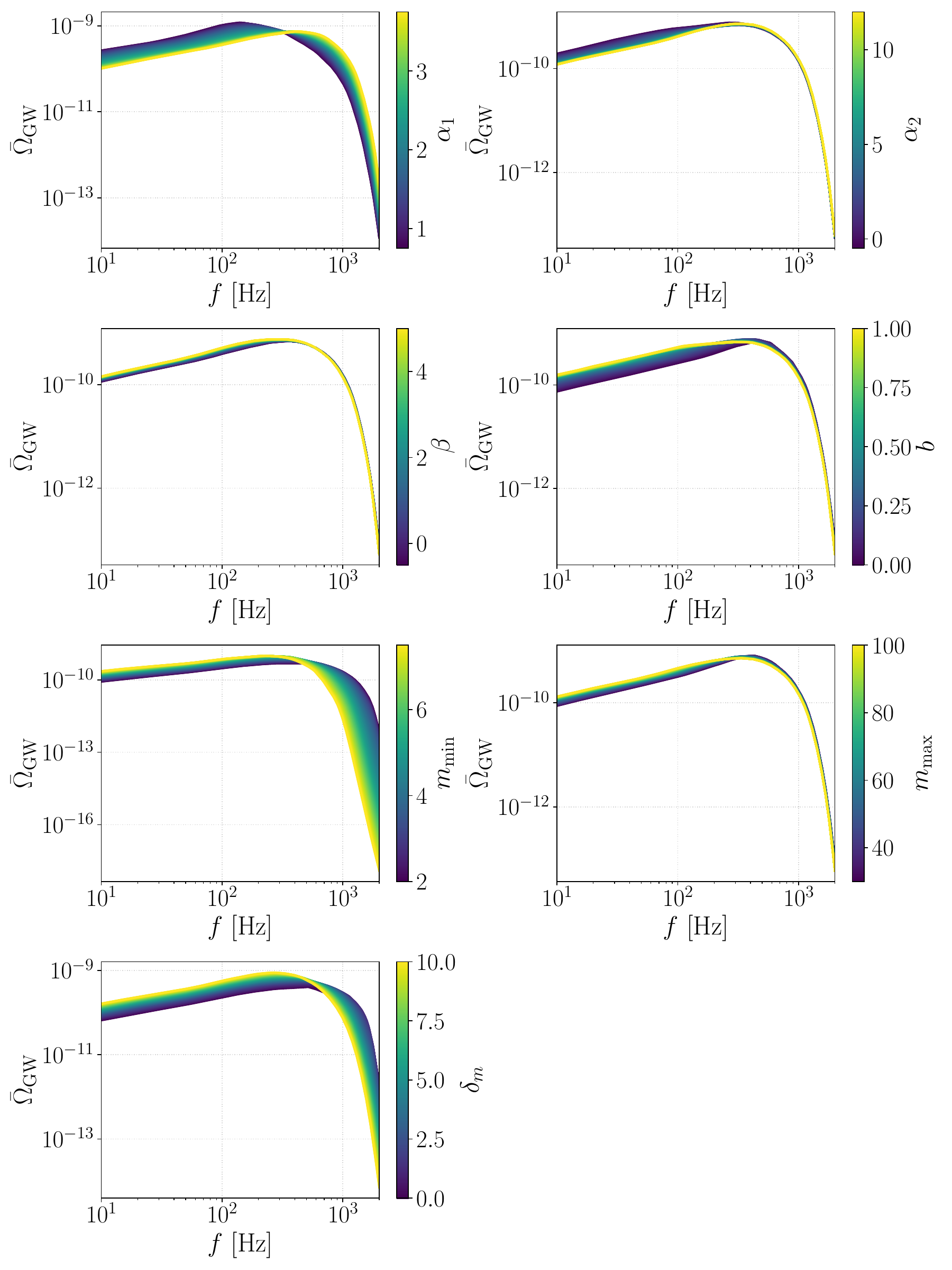}
     \caption{Impact of mass parameters on the predicted SGWB spectrum for the BPL+MD model. Each panel illustrates the SGWB spectrum $\bar \Omega_{\text{GW}}(f)$ as a function of frequency $f$. The varying parameter ($\alpha_1$, $\alpha_2$, $\beta$, $b$, $m_{\min}$, $m_{\max}$ and $\delta_m$,) is indicated next to the color bar in each panel, with different color curves representing different values. All other parameters are kept fixed at their reference values. We can more clearly observe the impact of $\alpha_1$, which is analogous to the $\alpha$ parameter discussed for the PP+MD model in Fig.~\ref{fig:GWB_variation_plpp_mass}, since the range of variation is greater. Furthermore, these plots again highlight the degeneracy between $m_{\min}$ and $\delta_m$.
     }
     \label{fig:GWB_variation_bpl_mass}
 \end{figure}

The impact of varying the redshift parameters assuming the PP+MD model is presented in Fig.~\ref{fig:GWB_variation_plpp_redshift}, while the impact of varying mass parameters is in Fig.~\ref{fig:GWB_variation_plpp_mass}. %
The impact of varying the redshift parameters assuming the BPL+MD model is presented in Fig.~\ref{fig:GWB_variation_bpl_redshift}, while the impact of varying mass parameters is in Fig.~\ref{fig:GWB_variation_bpl_mass}. %

To summarize which parameters were fixed, marginalized, or constrained in each inference run corresponding to a corner plot, we provide Table~\ref{summary_params}.
In both models, the redshift parameters which have a visible impact on the $\Omega_{\rm GW}$ spectrum (within the tested priors) are $R_0$, $\gamma$, and $z_{\rm peak}$. %

In the PP mass model, dominant parameters are $\alpha$ and $\Lambda_{\rm peak}$, as well as $m_{\rm min}$ and $\delta_m$, which are degenerate. %
Parameters which characterise the location and size of the peak ($\mu_{\rm peak}$ and $\sigma_{\rm peak}$) do not have significant impact on $\Omega_{\rm GW}$, most probably due to the fact that the fraction of black holes expected in the peak is small for the entire set ($\lambda_{\rm peak}\lesssim 0.2$). %
The secondary mass parameter $\beta$ and the maximum mass cut-off $m_{\rm max}$ are also sub-dominant.

In the BPL mass model, dominant parameters are $\alpha_1$, $\alpha_2$, and $b$, as well as $m_{\rm min}$ and $\delta_m$, which are degenerate (for both mass models) as discussed in Sec.~\ref{sec:bpl_md_article}. %
The $\beta$ and the $m_{\rm max}$ parameters are sub-dominant, as in the PP case.

\begin{table}
\caption{Summary of the PLPP+MD and BPL+MD model hyperparameters.}
\begin{minipage}{0.48\textwidth}
\begin{indented}
\item[]\begin{tabular}{@{}ll}
\br
\textbf{PP+MD} \\
\mr
$\alpha$ & 2.66 \\
$\beta$ & 0.81 \\
$\delta_m$ & 6.28 \\
$\lambda_{\mathrm{peak}}$ & 0.04 \\
$m_{\mathrm{max}}$ & 89.53 \\
$m_{\mathrm{min}}$ & 3.71 \\
$\mu_{\mathrm{peak}}$ & 35.78 \\
$\sigma_{\mathrm{peak}}$ & 2.89 \\
$\mathcal{R}_0$ & 17.36 \\
$\gamma$ & 2.08 \\
$\kappa$ & 6.78 \\
$z_{\mathrm{peak}}$ & 1.52 \\
\br
\end{tabular}
\end{indented}
\end{minipage}
\hfill
\begin{minipage}{0.48\textwidth}
\begin{indented}
\item[]\begin{tabular}{@{}ll}
\br
\textbf{BPL+MD} \\
\mr
$\alpha_1$ & 2.66 \\
$\alpha_2$ & 6.3 \\
$\beta$ & 0.81 \\
$b$ & 0.38 \\
$m_{\mathrm{min}}$ & 3.71 \\
$m_{\mathrm{max}}$ & 89.53 \\
$\delta_m$ & 6.28 \\
$\mathcal{R}_0$ & 17.36 \\
$\gamma$ & 2.08 \\
$\kappa$ & 6.78 \\
$z_{\mathrm{peak}}$ & 1.52 \\
\br
\end{tabular}
\end{indented}
\label{tab:bpl_params}
\end{minipage}
\label{tab:params_variations}
\end{table}

\begin{table}
\caption{\label{summary_params}Summary of parameters for each inference run. Constrained parameters are shown in the corner plots, marginalized parameters were sampled but not shown, and fixed parameters were held at injected values.}
\vspace{0.3cm}
\small
\centering
\begin{tabular}{{@{}llll}}
\br
\textbf{Inference Run} & \textbf{Constrained} & \textbf{Marginalized} & \textbf{Fixed} \\
\mr
MD (Fig.~\ref{fig:inference_redshift_md}) & $\mathcal{R}_0$, $\gamma$, $\kappa$, $z_{\rm peak}$ & $\alpha$, $\beta$, $m_{\rm min}$, $m_{\rm max}$,  & — \\ & & $\delta_m$, $\lambda_{\rm peak}$, $\mu_{\rm peak}$, $\sigma_{\rm peak}$ &  \\
\mr
PP+MD (Fig.~\ref{fig:inference_plpp}) & $\alpha$, $\lambda_{\rm peak}$, $\mathcal{R}_0$, $\gamma$ & — & $\beta$, $m_{\rm min}$, $m_{\rm max}$, $\delta_m$ \\ & & & $\mu_{\rm peak}$, $\sigma_{\rm peak}$, , $\kappa$, $z_{\rm peak}$ \\
\mr
BPL+MD (Fig.~\ref{fig:inference_bpl}) & $\alpha_1$, $b$, $m_{\rm min}$, $\delta_m$ & — & $\alpha_2$, $\beta$, $m_{\mathrm{max}}$ \\ & & & $\mathcal{R}_0$, $\gamma$, $\kappa$, $z_{\rm peak}$ \\
\br
\end{tabular}
\end{table}

\section{Waveform model sensitivity check} \label{appendix:waveform_model_sensitivity_check}

\begin{figure}
    \centering
    \includegraphics[width=0.95\linewidth]{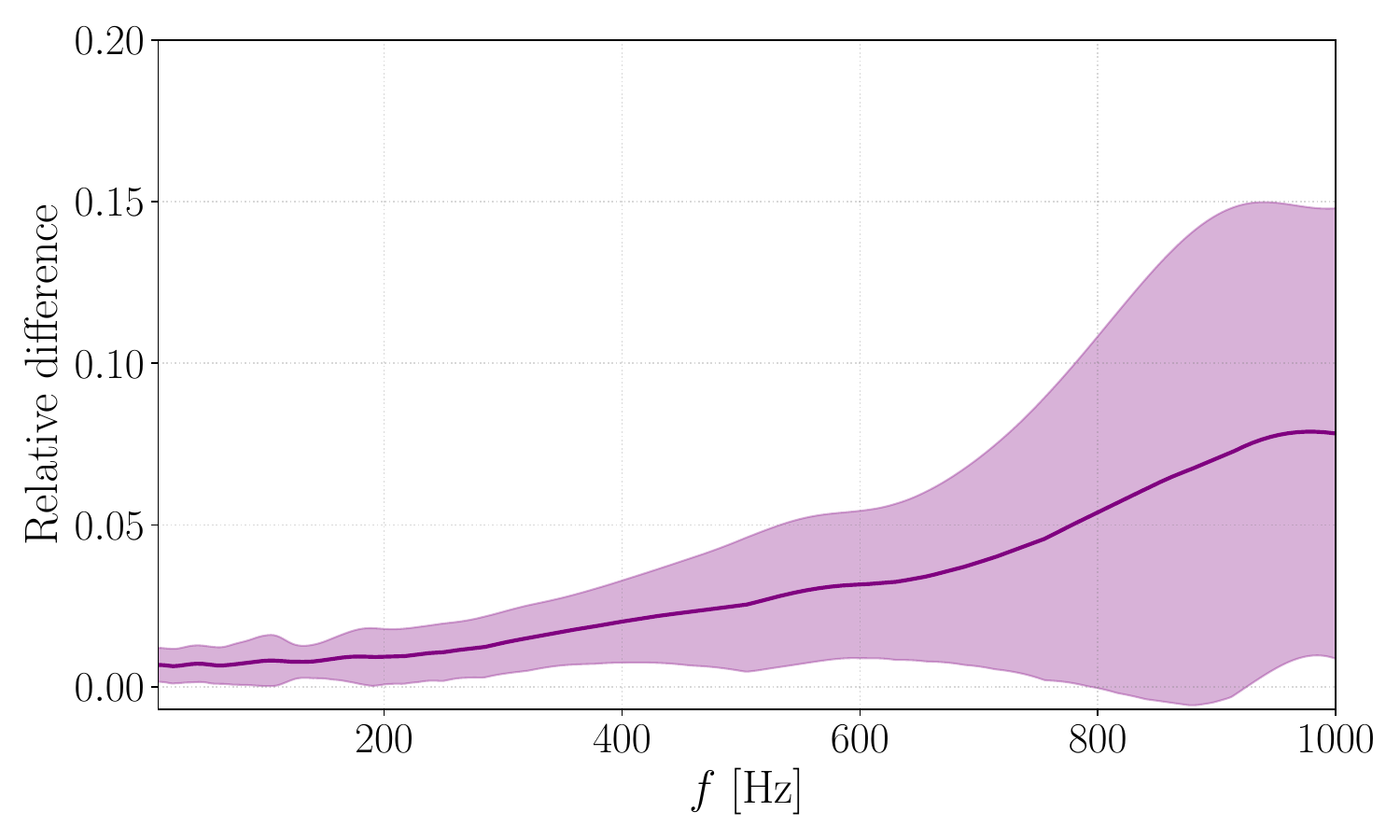} 
    \caption{Relative difference between SGWB spectra obtained with \textsc{IMRPhenomD} and \textsc{IMRPhenomXAS} as a function of frequency. The deviations remain below $5\%$ across the frequency band relevant for inference.}
    \label{fig:waveform_sensitivity}
\end{figure}

To verify that the choice of waveform approximant does not significantly affect our results, we compare the SGWB spectra computed with \textsc{IMRPhenomD} and \textsc{IMRPhenomXAS} \cite{2020PhRvD.102f4001P}.  
Figure~\ref{fig:waveform_sensitivity} shows the frequency-dependent relative difference. The comparison shows that the deviations remain below $5\%$ across the frequency band relevant for inference, indicating that our results are robust against the choice of waveform model.

\section{Sensitivity of non-colocated detectors to the GWB}\label{appendix:noncolocated_detectors}

\begin{figure}
    \centering
    \includegraphics[width=0.95\textwidth]{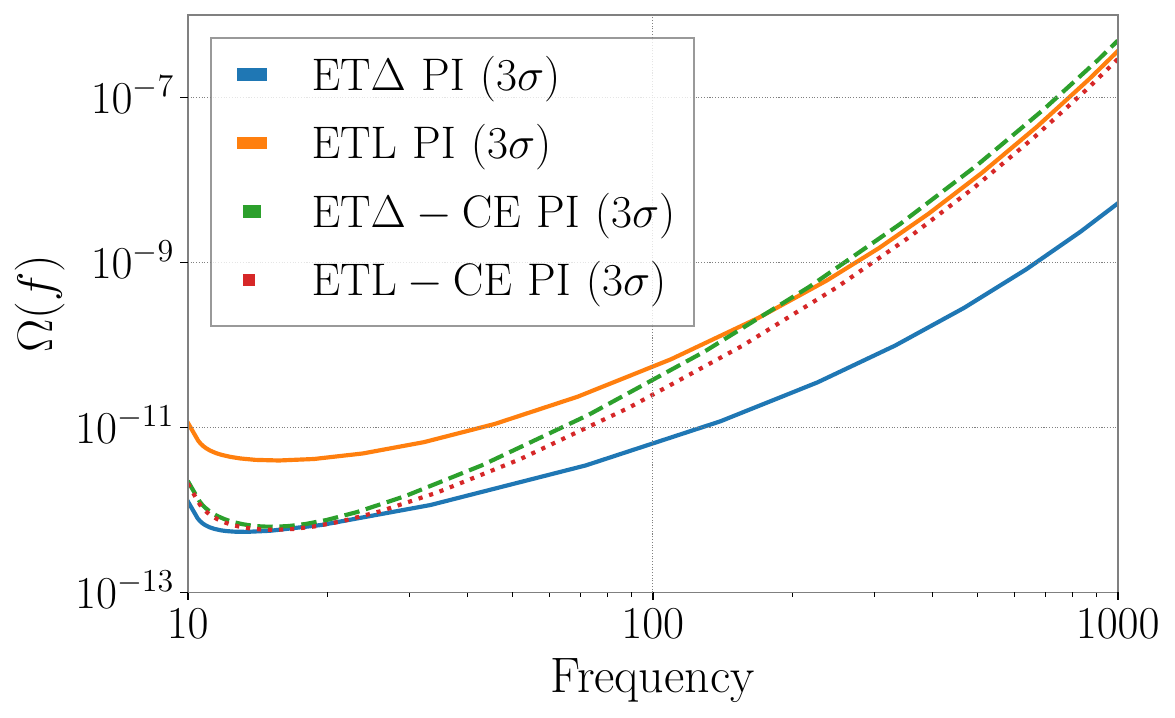}
    \caption{Power-law integrated sensitivity curves (PI curves) for different detector configurations assuming $3\sigma$ measurements. Co-located, triangular ET (blue); two non-colocated L-shaped ET instruments (orange); a triangular ET (in Europe) correlated with a CE instrument (in the USA) (green); and a single L-shaped ET with a CE instrument (red). We assume standard locations and orientations for these instruments, as used throughout this work. Comparing the colocated triangluar ET case with the non-colocated L-shaped ET we find that the sensitivity varies between a factor of 5 at low frequency ($<100$ Hz) to a factor of 100 at high frequency ($>1000$ Hz).}
    \label{fig:non_colocated_example}
\end{figure}

To quantify the impact of non-colocated detectors in our analysis, we compare the power-law integrated sensitivity curves (PI curves)~\cite{2013PhRvD..88l4032T} for different configurations, similarly to~\cite{2023JCAP...07..068B}. %
The error on the measurement will scale according to the PI curve, which is defined as the locus where an envelope of power-law signals is tangent to the detector sensitivity, assuming a certain detection statistic (i.e., signal-to-noise ratio). In Fig.~\ref{fig:non_colocated_example}, we show PI curves assuming $3\sigma$ measurements for four configurations: co-located, triangular ET; two non-colocated L-shaped ET instruments; a triangular ET (in Europe) correlated with a CE instrument (in the USA); and a single L-shaped ET with a CE instrument. We assume standard locations and orientations for these instruments, as used throughout this work. For the L-shaped ET detector pair, we use conventions as presented in~\cite{2021CQGra..38q5014B}. Non-colocated detector pairs have lower sensitivity to GWBs due to the decoherence induced by the overlap reduction function~\cite{2013PhRvD..88l4032T}. In particular, comparing the colocated triangluar ET case with the non-colocated L-shaped ET we find that the sensitivity varies between a factor of 5 at low frequency ($<100$ Hz) to a factor of 100 at high frequency ($>1000$ Hz). On the other hand, cross-correlating an ET detector with CE appears to be comparable to a triangular ET configuration at low frequency, due to the increased sensitivity of CE. As the analysis presented here is dominated by the intrinsic variance of the signal, and not the detector noise, it is not immediately clear how this would impact parameter inference.

\end{document}